\documentclass[letterpaper,12pt]{article}
\usepackage[utf8]{inputenc}
\usepackage{tabularx} 
\usepackage{amsmath}  
\usepackage{bm}
\usepackage{graphicx} 
\usepackage[margin=1in,letterpaper]{geometry} 
\usepackage{cite} 
\usepackage[final]{hyperref} 
\hypersetup{
	colorlinks=true,       
	linkcolor=blue,        
	citecolor=blue,        
	filecolor=magenta,     
	urlcolor=blue
}

\usepackage{setspace} 
\usepackage{subcaption}
\usepackage{booktabs}
\usepackage{nicefrac}
\usepackage[multiple]{footmisc}
\usepackage{multirow}
\usepackage{xargs}             
\usepackage[colorinlistoftodos,prependcaption,textsize=tiny]{todonotes}
\newcommandx{\unsure}[2][1=]{\todo[linecolor=red,backgroundcolor=red!25,bordercolor=red,#1]{#2}}
\newcommandx{\change}[2][1=]{\todo[linecolor=blue,backgroundcolor=blue!25,bordercolor=blue,#1]{#2}}
\newcommandx{\info}[2][1=]{\todo[linecolor=OliveGreen,backgroundcolor=OliveGreen!25,bordercolor=OliveGreen,#1]{#2}}
\newcommandx{\improvement}[2][1=]{\todo[linecolor=Plum,backgroundcolor=Plum!25,bordercolor=Plum,#1]{#2}}
\newcommandx{\thiswillnotshow}[2][1=]{\todo[disable,#1]{#2}}

\graphicspath{{Figures/}}

\usepackage{tikz}
\usepackage{tikz-network}
\usetikzlibrary{arrows,decorations.pathmorphing,backgrounds,positioning,fit, calc, shapes, patterns, patterns.meta}

\tikzset{My Arrow Style/.style={single arrow, fill=black!50, anchor=base, align=center,text width=.7\textwidth}}
\newcommand{\MyArrow}[2][]{\tikz[baseline] \node [My Arrow Style,#1] {#2};}

\newlength{\hatchspread}
\newlength{\hatchthickness}
\newlength{\hatchshift}
\newcommand{\hatchcolor}{}
\tikzset{hatchspread/.code={\setlength{\hatchspread}{#1}},
         hatchthickness/.code={\setlength{\hatchthickness}{#1}},
         hatchshift/.code={\setlength{\hatchshift}{#1}},
         hatchcolor/.code={\renewcommand{\hatchcolor}{#1}}}
\tikzset{hatchspread=3pt,
         hatchthickness=0.4pt,
         hatchshift=0pt,
         hatchcolor=black}
\pgfdeclarepatternformonly[\hatchspread,\hatchthickness,\hatchshift,\hatchcolor]
   {custom north west lines}
   {\pgfqpoint{\dimexpr-2\hatchthickness}{\dimexpr-2\hatchthickness}}
   {\pgfqpoint{\dimexpr\hatchspread+2\hatchthickness}{\dimexpr\hatchspread+2\hatchthickness}}
   {\pgfqpoint{\dimexpr\hatchspread}{\dimexpr\hatchspread}}
   {
    \pgfsetlinewidth{\hatchthickness}
    \pgfpathmoveto{\pgfqpoint{0pt}{\dimexpr\hatchspread+\hatchshift}}
    \pgfpathlineto{\pgfqpoint{\dimexpr\hatchspread+0.15pt+\hatchshift}{-0.15pt}}
    \ifdim \hatchshift > 0pt
      \pgfpathmoveto{\pgfqpoint{0pt}{\hatchshift}}
      \pgfpathlineto{\pgfqpoint{\dimexpr0.15pt+\hatchshift}{-0.15pt}}
    \fi
    \pgfsetstrokecolor{\hatchcolor}
    \pgfusepath{stroke}
   }

\usepackage{array}
\usepackage{authblk}
\usepackage{longtable}
\usepackage{makecell}
\usepackage{dcolumn}
\newcolumntype{d}[1]{D{.}{.}{#1}}

\usepackage[listings,many]{tcolorbox}

\definecolor{mygrey}{RGB}{64,64,64}

\tcbset{listing engine=listings}
\tcbset{mystyle/.style={
  breakable,
  enhanced,
  outer arc=0pt,
  arc=0pt,
  colframe=mygrey,
  colback=white,
  attach boxed title to top left,
  boxed title style={
  colback=mygrey,
  outer arc=0pt,
  arc=0pt,
  top=3pt,
  bottom=3pt,
  },
  fonttitle=\sffamily
  }
}

\newtcolorbox[auto counter,number within=section]{mybox}[2][]{
  mystyle,
  floatplacement=htbp,
  float,
  title=Summary~\thetcbcounter,
  overlay unbroken and first={
      \path
        let
        \p1=(title.north east),
        \p2=(frame.north east)
        in
        node[anchor=west,font=\sffamily,text width=\x2-\x1]
        at (title.east) {#1};
  },
  label=#2
}

\begin{document}

\title{Assortative Mixing in Weighted Directed Networks}
\author{U. Pigorsch and M. Sabek}

\affil{\small Schumpeter School of Business and Economics, University of Wuppertal,
Gaußstraße 20, D-42119 Wuppertal, Germany. Email: sabek@uni-wuppertal.de, pigorsch@uni-wuppertal.de}

\date{\today}

\maketitle

\onehalfspacing

\begin{abstract}
A network's assortativity is the tendency of vertices to bond with others based on similarities, usually excess vertex degree. In this paper we consider assortativity in weighted networks, both directed and undirected. To this end, we propose to consider excess vertex strength, rather than excess degree, and show, that assortativity in weighted networks can be broken down into two mechanisms, which we refer to as the connection effect and the amplification effect. To capture these effects we introduce a generalised assortativity coefficient. This new coefficient allows for a more detailed interpretation and assessment of assortativity in weighted networks. Furthermore, we propose a procedure to assess the statistical significance of assortativity using jackknife, bootstrap and rewiring techniques. The usefulness of our proposed generalised assortativity coefficient is demonstrated by an in-depth analysis of the assortativity structure of several weighted real-world networks.
\end{abstract}

\section{Introduction}

Assortativity is the tendency of a vertex to bond with another based on their similarity, with similarity being usually measured via vertex degree.
The most popular assortativity measure is the assortativity coefficient, which is defined as the Pearson correlation coefficient between the excess degrees of both ends of an edge.
This very popular assortativity measure has been originally proposed by Newman \cite{Newman2002,Newman2003} for unweighted and undirected networks and extended to directed networks by computing the correlation between the excess out- and in-degree of both ends of an edge. \cite{Piraveenan}, instead, suggest to compute the correlation between the excess out-degrees or between the excess in-degrees providing separate measures for out-assortativity and in-assortativity, and, as such, additional information with respect to the topology of the analysed network.

So far, as already pointed out in\, \cite{Noldus}, assortativity in weighted networks has been insufficiently studied, which is surprising as many real-world networks exhibit weighted edges.
In \cite{Leung} weighted, but undirected networks are considered.
In such networks, the assortativity coefficient by Newman \cite{Newman2002} can be computed, but it neglects important information about the intensity of the interaction between two vertices.
The authors therefore propose a weighted assortativity coefficient which takes edge weights into account.
However, as we will detail below, this weighted assortativity coefficient falls short in the sense that it does not consider the strengths of vertices, but focuses solely on their degrees.

In this paper we propose a more general coefficient of assortativity that nests the afore-mentioned assortativity measures as special cases, and that can be applied to (un)weighted, (un)directed networks. Moreover, we show that the use of this general coefficient enables us to determine the underlying assortativity structure in weighted networks more precisely. Furthermore, we propose a procedure to assess the statistical significance of assortativity using jackknife, bootstrap and rewiring techniques.

The remainder of the paper is structured as follows. Section 2 provides a review of the related literature and motivates the use of excess strength for the computation of assortativity. In Section 3 we introduce our generalised assortativity coefficient, elaborate on the importance of considering excess strength rather than total strength, and propose procedures for the interpretation and statistical assessment of assortativity in weighted networks. Section 4 illustrates the application and interpretation of assortativity in weighted real-world networks. Section 5 concludes.   

\section{Background and Related Literature}
The assortativity coefficient $r^{\text{N}}$ has been proposed by  Newman \cite{Newman2002} and is defined as the Pearson correlation coefficient between the excess degrees (sometimes: remaining degrees) of both ends of an edge. Excess or remaining degrees are defined to be one less than the ends' degrees, i.e. they are the degrees of the ends prior to the formation of the particular edge which is currently considered. The coefficient is obtained by computing
\begin{align}
	r^\text{N} = \frac{M^{-1}\sum_i j_ik_i-\big[M^{-1}\sum_i\frac{1}{2}(j_i+k_i)\big]^2}{M^{-1}\sum_i\frac{1}{2}(j_i^2+k_i^2)-\big[M^{-1}\sum_i\frac{1}{2}(j_i+k_i)\big]^2},
	\label{eq:r_N}
\end{align}
where $j_i$ and $k_i$ are the excess degrees of the ends $j$ and $k$ of edge $i$, where $i = 1, \hdots, M$, and $M$ is the number of edges in the network. Since $r^{\text{N}}$ is a correlation coefficient it lies in the range $-1 \leq r^{\text{N}} \leq 1$, and has the advantage that assortativity coefficients can be compared across different networks. The coefficient in Eq.\,(\ref{eq:r_N}) has been proposed for \textit{undirected} \textit{unweighted} networks.

An extension towards \textit{directed} \textit{unweighted} networks has also been proposed by Newman \cite{Newman2003}, who defines assortativity in directed networks as the correlation coefficient between the excess out-degree of the vertex that the $i$-th edge leads out of and the excess in-degree of the vertex that the $i$-th edge leads into.
In addition to this,\, \cite{Piraveenan} find it sensible to also consider both the correlation between (excess) out-degrees of both ends of an edge and the correlation between (excess) in-degrees of both ends of an edge.
Therefore, they propose alternative definitions for assortativity in directed networks, namely out-assortativity, which is the tendency of vertices to bond with others of similar out-degree as themselves and in-assortivity, which is the tendency of vertices to bond with others of similar in-degree as themselves. This results in four different variants of the assortativity coefficient in directed networks.

For our empirical analysis, we refer to these variants as the \textit{mode of assortativity} and denote by \textit{out-in} the assortativity coefficient of Newman \cite{Newman2003}, by \textit{out-out} and \textit{in-in} the out- and in-assortativity according to \cite{Piraveenan}, respectively, and lastly, by \textit{in-out} the correlation coefficient between the in-degree of the vertex that the $i$-th edge leads out of and the out-degree of the vertex that the $i$-th edge leads into, see also \cite{Piraveenan}.

The corresponding assortativity coefficient for \textit{directed} networks is given by:
\begin{align}
	r^\text{N}_d = \frac{\sum_i j^\prime_ik^\prime_i-M^{-1}\big(\sum_i j^\prime_i\big)\big(\sum_i k^\prime_i\big)}{\sqrt{\Big[\sum_i (j^\prime_i)^2 - M^{-1}\big(\sum_i j^\prime_i\big)^2\Big]\Big[\sum_i (k^\prime_i)^2 - M^{-1}\big(\sum_i k^\prime_i\big)^2\Big]}},
	\label{eq:r_N_d}
\end{align}
where this time $j^\prime_i$ and $k^\prime_i$ are the (excess) in- or out-degrees of ends $j$ and $k$ of the $i$-th edge, and $i = 1, \hdots, M$.

The coefficient in Eq.\,(\ref{eq:r_N_d}) has been introduced for \textit{directed} \textit{unweighted} networks. However, it is capable of handling \textit{undirected} \textit{unweighted} networks as well, if the network is slightly modified, i.e., by replacing each undirected edge by two directed ones that point in opposite directions, see \cite{Newman2003}. Indeed, the formulation in Eq.\,(\ref{eq:r_N}) is a simplification of the more general formulation in Eq.\,(\ref{eq:r_N_d}) that makes use of the property of symmetry of the adjacency matrix of an undirected network.

According to\, \cite{Noldus}, assortativity in weighted networks has been insufficiently studied, so far. One exception is\, \cite{Leung}, where the following extension of the assortativity coefficient towards \textit{undirected} \textit{weighted} networks is suggested
\begin{equation}
    r^\text{LC} = \frac{H^{-1}\sum_i \omega_i(j_ik_i)-\big[H^{-1}\sum_i \frac{1}{2} \omega_i (j_i+k_i)\big]^2}{H^{-1}\sum_i \frac{1}{2} \omega_i (j_i^2+k_i^2)-\big[H^{-1}\sum_i \frac{1}{2} \omega_i (j_i+k_i)\big]^2},
    \label{eq:r_lc}
\end{equation}
where, as in Eq.\,(\ref{eq:r_N}), $j_i$ and $k_i$ are the excess degrees of the ends $j$ and $k$ of edge $i$, $\omega_i$ denotes the weight of the $i$-th edge and $H=\sum_i \omega_i$ is the sum of edge weights where the sum is over all edges.
Obviously, if all edge weights equal one, i.e. the network is unweighted, the coefficient in Eq.\,(\ref{eq:r_lc}) reduces to the original assortativity coefficient in Eq.\,(\ref{eq:r_N}).

The underlying mechanism of this assortativity coefficient, can easily be illustrated.
For the ease of exposition and without loss of generality suppose integer-valued weights (as real-valued weights can be linearly mapped to integers with arbitrary precision, see \cite{Rubinov1}).\footnote{For example, $ w_\text{int} = \text{round}\big(w_\text{real}\cdot10^\text{precision}\big) $ is a mapping that linearly maps real-valued weights to integers with arbitrary precision.}
Then incorporating edge weights is equivalent to replacing each $\omega$-weighted edge by $\omega$ edges with weight one.
Thus, high-weighted edges amplify the impact of their connections and therefore contribute more to the overall assortativity.

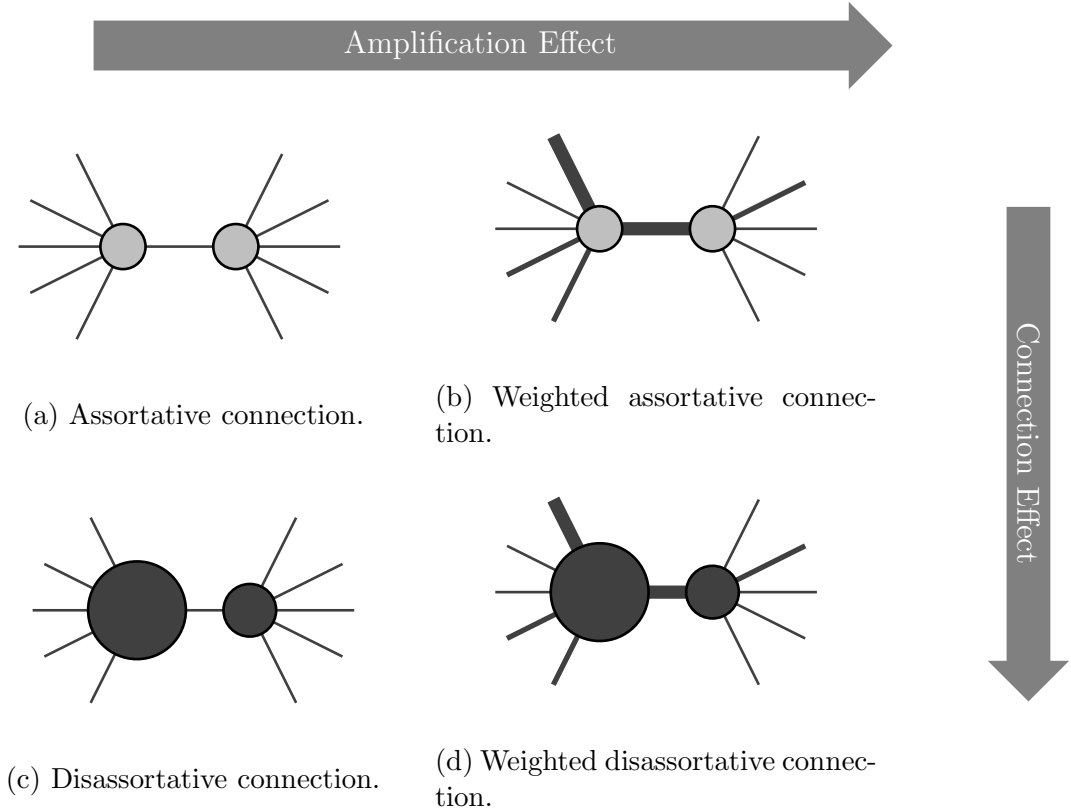
\begin{figure}[t]
	\centering
	\begin{subfigure}{\textwidth}
		\centering
		\MyArrow[align = center, text width = 10cm]{\color{white} Amplification Effect}
	\end{subfigure}%

	\begin{subfigure}{\textwidth}
		\begin{subfigure}{.89\textwidth}
			\centering
			\begin{subfigure}{0.4\textwidth}
				\begin{tikzpicture}[scale=1.5]
					\Vertex[color=lightgray]{A}
					\Vertex[x=1,color=lightgray]{B}
					\Vertex[x=1.5,y=1,Pseudo,color=gray]{C}
					\Vertex[x=2,y=.5,Pseudo,color=gray]{D}
					\Vertex[x=2,y=-.5, Pseudo,color=gray]{E}
					\Vertex[x=1.5,y=-1,Pseudo,color=gray]{F}
					\Vertex[x=-.5,y=1,Pseudo,color=gray]{G}
					\Vertex[x=-1,y=.5,Pseudo,color=gray]{H}
					\Vertex[x=-1,y=-.5,Pseudo,color=gray]{I}
					\Vertex[x=-.5,y=-1,Pseudo,color=gray]{J}
					\Vertex[x=2.125,y=0,Pseudo,color=gray]{K}
					\Vertex[x=-1.125,y=0,Pseudo,color=gray]{L}
					\Edge[lw=1pt](A)(B)
					\Edge[lw=1pt](B)(C)
					\Edge[lw=1pt](B)(D)
					\Edge[lw=1pt](B)(E)
					\Edge[lw=1pt](B)(F)
					\Edge[lw=1pt](A)(G)
					\Edge[lw=1pt](A)(H)
					\Edge[lw=1pt](A)(I)
					\Edge[lw=1pt](A)(J)
					\Edge[lw=1pt](B)(K)
					\Edge[lw=1pt](A)(L)
				\end{tikzpicture}
				\caption{Assortative connection.}
				\label{fig:a}
			\end{subfigure}%
			~
			\begin{subfigure}{0.4\textwidth}
				\centering
				\begin{tikzpicture}[scale=1.5]
				\Vertex[color=lightgray]{A}
				\Vertex[x=1,color=lightgray]{B}
				\Vertex[x=1.5,y=1,Pseudo,color=gray]{C}
				\Vertex[x=2,y=.5,Pseudo,color=gray]{D}
				\Vertex[x=2,y=-.5, Pseudo,color=gray]{E}
				\Vertex[x=1.5,y=-1,Pseudo,color=gray]{F}
				\Vertex[x=-.5,y=1,Pseudo,color=gray]{G}
				\Vertex[x=-1,y=.5,Pseudo,color=gray]{H}
				\Vertex[x=-1,y=-.5,Pseudo,color=gray]{I}
				\Vertex[x=-.5,y=-1,Pseudo,color=gray]{J}
				\Vertex[x=2.125,y=0,Pseudo,color=gray]{K}
				\Vertex[x=-1.125,y=0,Pseudo,color=gray]{L}
				\Edge[lw=5pt](A)(B)
				\Edge[lw=1pt](B)(C)
				\Edge[lw=2pt](B)(D)
				\Edge[lw=1pt](B)(E)
				\Edge[lw=1pt](B)(F)
				\Edge[lw=5pt](A)(G)
				\Edge[lw=1pt](A)(H)
				\Edge[lw=2pt](A)(I)
				\Edge[lw=2pt](A)(J)
				\Edge[lw=1pt](B)(K)
				\Edge[lw=1pt](A)(L)
				\end{tikzpicture}
				\caption{Weighted assortative connection.}
				\label{fig:b}
			\end{subfigure}%

			\begin{subfigure}{0.4\textwidth}
				\centering
				\begin{tikzpicture}[scale=1.5]
				\Vertex[color=darkgray, size=1.3]{A}
				\Vertex[x=1,color=darkgray, size=0.7]{B}
				\Vertex[x=1.5,y=1,Pseudo,color=gray]{C}
				\Vertex[x=2,y=.5,Pseudo,color=gray]{D}
				\Vertex[x=2,y=-.5, Pseudo,color=gray]{E}
				\Vertex[x=1.5,y=-1,Pseudo,color=gray]{F}
				\Vertex[x=-.5,y=1,Pseudo,color=gray]{G}
				\Vertex[x=-1,y=.5,Pseudo,color=gray]{H}
				\Vertex[x=-1,y=-.5,Pseudo,color=gray]{I}
				\Vertex[x=-.5,y=-1,Pseudo,color=gray]{J}
				\Vertex[x=2.125,y=0,Pseudo,color=gray]{K}
				\Vertex[x=-1.125,y=0,Pseudo,color=gray]{L}
				\Edge[lw=1pt](A)(B)
				\Edge[lw=1pt](B)(C)
				\Edge[lw=1pt](B)(D)
				\Edge[lw=1pt](B)(E)
				\Edge[lw=1pt](B)(F)
				\Edge[lw=1pt](A)(G)
				\Edge[lw=1pt](A)(H)
				\Edge[lw=1pt](A)(I)
				\Edge[lw=1pt](A)(J)
				\Edge[lw=1pt](B)(K)
				\Edge[lw=1pt](A)(L)
				\end{tikzpicture}
				\caption{Disassortative connection.}
				\label{fig:c}
			\end{subfigure}%
			~
			\begin{subfigure}{0.4\textwidth}
				\centering
				\begin{tikzpicture}[scale=1.5]
				\Vertex[color=darkgray, size=1.3]{A}
				\Vertex[x=1,color=darkgray, size=0.7]{B}
				\Vertex[x=1.5,y=1,Pseudo,color=gray]{C}
				\Vertex[x=2,y=.5,Pseudo,color=gray]{D}
				\Vertex[x=2,y=-.5, Pseudo,color=gray]{E}
				\Vertex[x=1.5,y=-1,Pseudo,color=gray]{F}
				\Vertex[x=-.5,y=1,Pseudo,color=gray]{G}
				\Vertex[x=-1,y=.5,Pseudo,color=gray]{H}
				\Vertex[x=-1,y=-.5,Pseudo,color=gray]{I}
				\Vertex[x=-.5,y=-1,Pseudo,color=gray]{J}
				\Vertex[x=2.125,y=0,Pseudo,color=gray]{K}
				\Vertex[x=-1.125,y=0,Pseudo,color=gray]{L}
				\Edge[lw=5pt](A)(B)
				\Edge[lw=1pt](B)(C)
				\Edge[lw=2pt](B)(D)
				\Edge[lw=1pt](B)(E)
				\Edge[lw=1pt](B)(F)
				\Edge[lw=5pt](A)(G)
				\Edge[lw=1pt](A)(H)
				\Edge[lw=2pt](A)(I)
				\Edge[lw=2pt](A)(J)
				\Edge[lw=1pt](B)(K)
				\Edge[lw=1pt](A)(L)
				\end{tikzpicture}
				\caption{Weighted disassortative connection.}
				\label{fig:d}
			\end{subfigure}
		\end{subfigure}
		\begin{subfigure}{.09\textwidth}
			\MyArrow[align = center, text width = 6cm, rotate=-90]{\color{white} Connection Effect}
		\end{subfigure}
	\end{subfigure}
	\caption{Illustration of  the connection effect  and the amplification effect. All figures show the same excerpt of a simple example network. The focus lies on the connection between the depicted vertices. The size of a vertex is drawn proportional to the corresponding vertex value. In figures (a) and (b) excess degrees are used as vertex values (lightgray) whereas in figures (c) and (d) excess strengths are used (darkgray). The widths of edges are proportional to their respective weights in figures (b) and (d) whereas in (a) and (c) they are constant (equal weights). Arrows indicate the direction in which the respective mechanisms operate.}
	\label{fig:mechanisms}
\end{figure}

Although this is a reasonable approach, we instead propose a generalisation of assortativity to weighted networks that is based on the correlation between the \textit{excess strengths} of both ends of an edge. Considering excess strength is quite intuitive here, as vertex degree generalises to vertex strength in weighted networks, see \cite{Barrat}, who defines the strength of a vertex $u$ to be the total weight of its connections, i.e. $s_u=\sum_{v\in V} w_{uv}$ where $V$ is the vertex set and $w_{uv}$ is the weight of the edge between $u$ and $v$.

In fact, note that the emergence of assortativity in a weighted network consists of two mechanisms, see Fig.\,\ref{fig:mechanisms}. The first one is the just mentioned \textit{amplification effect}, which occurs if a connection is considered according to the respective edge weight when computing the correlation between the vertex values. For example, comparisons of Fig.\,\ref{fig:a} with Fig.\,\ref{fig:b} and Fig.\,\ref{fig:c} with Fig.\,\ref{fig:d}, respectively, show that an existing assortative (or disassortative) connection is amplified when edge weights are introduced. The second one is the \textit{connection effect}, which occurs if, instead of unweighted vertex values (e.g. excess degrees), weighted vertex values are considered (e.g. excess strengths). More precisely, consider two arbitrary adjacent vertices and suppose they have the same degrees but different strengths, as depicted in Fig.\,\ref{fig:mechanisms}.
The connection between them, weighted or not, is assortative if degrees are used as vertex values, but is disassortative if strengths are used, compare, for example, Fig.\,\ref{fig:a} with Fig.\,\ref{fig:c} or Fig.\,\ref{fig:b} with Fig.\,\ref{fig:d}. The connection effect might also occur vice versa, for example, if two adjacent vertices have different degrees but similar strengths. Noteworthy, the connection effect is ignored in the definition of the assortativity coefficient of Eq.\,(\ref{eq:r_lc}). In the following section we propose a generalized assortativity coefficient that incorporates both of these effects.

\section{A Generalised Assortativity Coefficient}
To account for both effects, we include vertex strength in addition to vertex degree into our assortativity coefficient, which is defined in Section\,\ref{section_definition}. In Section\,\ref{section_excess_strengths}, we elaborate on the importance of considering excess strength rather than total strength, a distinction that is rarely made explicitly in the context of vertex degree in the existing literature. Moreover, our proposed generalized assortativity coefficient nests four different assortativity coefficients. We suggest to compute and interpret all of them, as their comparison provides new insights on the assortativity structure of weighted networks. This is detailed in Section\,\ref{section_procedure}. We further supplement the analysis by proposing a procedure for assessing both, the statistical significance of the four assortativity coefficients, as well as whether the observed assortativity structure has social, organizational originis or has been randomly generated, see Section\,\ref{section_significance}.  

\subsection{Definition of the Generalised Assortativity Coefficient} \label{section_definition}
In the following, we introduce our generalised weighted assortativity coefficient, that takes the amplification effect as well as the connection effect into account.
To this end, let $s^\prime_u=\sum_{v\in V} w_{uv}^\alpha$, $\alpha \in \{0,1\}$, be a modified version of vertex strength.
Clearly, if $\alpha = 1$ then $s^\prime_u = s_u$, whereas for $\alpha = 0$ it reduces to ordinary vertex degree.
Our generalised weighted assortativity coefficient is then defined as
\begin{equation}
    r^\omega_{(\alpha,\beta)} = \frac{\sum_i \omega^\beta_il_im_i-\Omega^{-1} \big(\sum_i \omega^\beta_il_i\big) \big(\sum_i \omega^\beta_im_i\big)}{
    \sqrt{
        \bigg[
            \sum_i \big(\omega^\beta_il_i^2\big)-\Omega^{-1}\big(\sum_i \omega^\beta_il_i\big)^2
        \bigg]
        \bigg[
            \sum_i \big(\omega^\beta_im_i^2\big)-\Omega^{-1}\big(\sum_i \omega^\beta_im_i\big)^2
        \bigg]
    }
},\label{eq:r_new}
\end{equation}
where $l_i$ and $m_i$ are the excess (in- or out-) strengths of the ends $l$ and $m$ of edge $i$.
For example, $l_i = s^\prime_l - \omega^\alpha_i$ is the excess strength of end $l$ of edge $i$.
Furthermore, $\Omega=\sum_i \omega^\beta_i$ with $\beta \in \{0,1\}$.
Obviously, if $\beta=1$ then $\Omega = H$, whereas for $\beta=0$ it reduces to the total number of edges in the network, i.e. $\Omega = M$. The generalisation is achieved by introducing $\alpha$ and $\beta$, which account for the two different mechanisms, i e. the connection effect and the edge amplification effect, respectively.
As such the previous definitions of assortativity are nested as special cases, in particular $r^\omega_{(\alpha = 0, \beta = 1)} = r^\text{LC}$ and $r^\omega_{(\alpha = 0, \beta = 0)} = r^\text{N}$.

In contrast to\, \cite{Leung}, whose formulation of $r^\text{LC}$ is based on the formulation of $r^\text{N}$ introduced by Newman\,\cite{Newman2002}, our coefficient $r^\omega_{(\alpha, \beta)}$ is based on the formulation $r^\text{N}_d$ of \,\cite{Newman2003,Farine},
and, thus, capable of handling \textit{directed} (\textit{weighted}) networks as well as \textit{undirected} (\textit{weighted}) networks by replacing, as before, each undirected edge by two directed ones that point in opposite directions.

Our assortativity coefficients $ r^\omega_{(\alpha,\beta)} $ is also inline with the definition of the weighted correlation coefficient, see \cite{Costa}, which is given by
\begin{equation}
	r_w = \frac{\sum_i w_iX_iY_i-\sum_i w_iX_i \sum_i w_iY_i}{\sqrt{\bigg(\sum_i w_iX_i^2-(\sum_i w_iX_i)^2\bigg)\bigg(\sum_i w_iY_i^2-(\sum_i w_i Y_i)^2\bigg)}},
	\label{eq:weighted_correlation}
\end{equation}
where the sums are over observations $i$, $X_i$ and $Y_i$ are the pair of values of variables $X$ and $Y$ corresponding to the $i$-th observation, $w_i$ is the weight attributed to this observation and $\sum_{i} w_i = 1$. Furthermore, if all weights $w_i$ are equal they cancel out and Eq.\,(\ref{eq:weighted_correlation}) reduces to the usual formula for the Pearson correlation coefficient, i.e. the unweighted correlation coefficient.
Defining variables $X_i = l_i$ and $Y_i = m_i$ and weights $w_i = \omega^\beta_i$, it becomes immediately clear that Eq.\,(\ref{eq:r_new}) and Eq.\,(\ref{eq:weighted_correlation}) are equivalent.\footnote{Where the additional requirement that $\sum_i \omega^\beta_i = \Omega = 1$ can be met without loss of generality by a suitable remapping of the observed edge weights.}

To summarize, assortativity in weighted networks is not unambiguously defined. In fact, there are four different ways edge weights can be treated, resulting in four different versions of the assortativity coefficient. First of all, if present edge weights are neglected, assortativity can be measured as the correlation between the excess degrees of both ends of an edge, $r^\omega_{(0,0)}$, which is the classical definition of assortativity introduced by Newman\,\cite{Newman2002}, henceforth referred to as the benchmark assortativity coefficient, also cf. Fig.\,\ref{fig:a}.\,\cite{Leung} suggest to measure assortativity by the weighted correlation between the excess degrees of both ends of an edge, $r^\omega_{(0,1)}$ cf. Fig.\,\ref{fig:b}, however, this corresponds to only partly considering edge weights, since the vertex values are still unweighted.
The remaining two versions of the assortativity coefficient have not been considered in the literature so far, and are both based on excess vertex strength, i.e. $\alpha=1$, rather than excess vertex degree. In particular, we can either partly incorporate edge weights, this time, by computing the unweighted correlation of weighted vertex values, i.e. excess strength, resulting in the assortativity coefficient $r^\omega_{(1,0)}$, or fully incorporate edge weight, i.e. $\alpha=1$ and $\beta=1$, by computing the weighted correlation between the excess strengths of both ends of an edge, cf. Fig.\,\ref{fig:c} and Fig.\,\ref{fig:d}. We denote the latter by $r^\omega_{(1,1)}$ and refer to it as the generalised assortativity coefficient.

When analysing the assortativity structure of a real weighted network, we suggest to focus on both, the generalised assortativity coefficient and the benchmark coefficient, i.e. to fully consider edge weights or to neglect them entirely.
For example, if the interest is exclusively on the binary network edges, it might be reasonable to neglect edge weights, and to focus on the benchmark assortativity coefficient.
However, in many cases edge weights provide additional information, which in turn can be fully explored using the generalised assortativity coefficient. In contrast, focusing exclusively on assortativity coefficients that only partially consider edge weights, falls short, as each includes just one of the two effects of edge weights. Nevertheless, as we will detail in Section \ref{section_procedure}, using them as supplementary measures allows to draw more distinct conclusions about the assortativity structure.

\subsection{Excess (Out- or In-) Strengths in Directed Weighted Networks}\label{section_excess_strengths}
The existing literature on the measurement of assortativity rarely explicitly addresses whether total degress or excess degrees is used; rather, it is oftentimes just referred to ``degree''. However, Newman\,\cite{Newman2002} defines the assortativity coefficient for undirected and unweighted networks to be the correlation coefficient between the excess degrees rather than the total degrees of both ends of an edge.
The reason for this is that a vertex's tendency to bond with another one is based on the degree it has prior to forming the particular edge, i.e., its own excess degree as well as the other vertex's excess degree, cf. \cite{Noldus}.
Consequently, using \textit{excess} strengths in case of weighted networks is the obvious choice.
We can only assume that the above mentioned imprecision is due to the fact that for \textit{unweighted} networks it makes no difference whether correlation is computed based on excess degrees or on total degrees, as both result in the same value of the assortativity coefficient. This holds as the excess degrees of the ends of an edge and the total degrees of the ends of an edge differ by a constant (i.e. by one) and the correlation between two random variables does not change if a constant is added or subtracted to either or both variables.
As opposed to this, in a \textit{weighted} network it indeed makes a difference whether excess or total strength is used.
The reason is that the excess strengths of the ends of an edge differ from total strengths of the ends of an edge by the weight of the particular edge and this difference is not constant.
Therefore, the resulting assortativity coefficient based on excess strengths will be different from the one based on total strengths.
More precisely, using total strengths rather than excess strenghts for computing the assortativity coefficient for a weighted network will lead to an overestimation towards the assortative direction, since high weighted edges necessarily connect vertices with high total strengths.
Hence, a connection between two vertices that might be disassortative will appear more assortative as the vertex values are artificially inflated by the weight of the edge that connects the two vertices when total strengths are used.
Thus, it is crucial to use excess strengths in order to properly determine the underlying assortativity structure of a network.\footnote{The Appendix contains an example that illustrates the consequences of using total strengths rather than excess strengths for the assortativity structure of the network considered in\,\cite{Yuan}.}

Based on this reasoning, we further recommend the following proper utilization of either excess or total strenghts (or degrees) when computing the assortativity coefficient for different modes of assortativity for \textit{directed} networks.
Excess strengths are used for both out- and in-strengths when the mode of assortativity is \textit{out-in}.
Excess out- and total out-strengths are used when the mode is \textit{out-out}, and total in- and excess in-strengths are used when the mode is \textit{in-in}.
For the mode \textit{in-out} the correlation between both total in- and out-strengths should be used.
To see this, suppose, for example, of interest is the out-assortativity of a directed \textit{weighted} network and consider a particular edge leading out of vertex $u$ and into vertex $v$, then, the out-strength of vertex $u$ is affected by the edge weight whereas the out-strength of vertex $v$ is not.
Hence, if we consider the out-strengths of the vertices that the particular edge connects prior to forming it, we get the excess out-strength for vertex $u$, which is its out-strength less the edge weight, and the excess out-strength for vertex $v$, which in this case equals its total out-strength.
For the other modes the reasoning is similar. Technically, the same holds true for directed \textit{unweighted} networks, however, the results are the same no matter if one computes the correlation between excess or total (in- or out-) degrees (or any combination), as before.

We have recently noticed that, in independent and concurrent research, the\,\cite{Yuan} have proposed a measure for assortativity in weighted networks similar to ours. Their measure, however, is based on the total strengths between the ends of an edge, which leads to misleading results as outlined above. Moreover, their paper focuses on the assortativity of theoretical network models, whereas a key contribution of our paper is the introduction of a procedure that allows for both, a more precise assessment and interpretation of the assortativity of weighted real-world networks, as well as its analysis with respect to the statistical significance of the network's assortativity.

\subsection{Procedure for Assessing and Interpretating a Network's Assortativity}\label{section_procedure}

In order to asses and interpret a network's assortativity, we suggest the following procedure: Firstly, compute $r^\omega_{(\alpha, \beta)}$ for all four parameter combinations $(\alpha, \beta)$, where $\alpha, \beta \in \{0,1\}$, i.e., we compute the benchmark assortativity coefficient $r^\omega_{(0,0)}$, the generalised assortativity coefficient $r^\omega_{(1,1)}$ as well as both supplementary measures $r^\omega_{(1,0)}$ and $r^\omega_{(0,1)}$.

The values of the benchmark assortativity coefficient as well as the generalised assortativity coefficient range between $-1$ and $1$ and give an indication of the underlying assortativity structure of the network with respect to the corresponding vertex values, which are degrees in case of $r^\omega_{(0,0)}$ and strenghts in case of $r^\omega_{(1,1)}$. Similar to the interpretation of the original assortativity coefficient, for both coefficients, positive values indicate an overall assortative structure of the network, and negative values indicate an overall disassortative structure of the network, for zero values of the coefficients the network is considered to be non-assortative.

Secondly, compare the benchmark and the generalised assortativity coefficient. The values of $r^\omega_{(0,0)}$ and $r^\omega_{(1,1)}$ might be similar in magnitude for some networks, for others they might differ.
Thus, a comparison of both values provides information on the impact of edge weights on the underlying assortativity structure of the network. For example, if $r^\omega_{(0,0)} > r^\omega_{(1,1)}$, then the consideration of edge weights leads to a decrease in assortativity or an increase in disassortativity of the network. In contrast, if $r^\omega_{(0,0)} < r^\omega_{(1,1)}$, the corresponding weighted network is more assortative than the network where edge weights are neglected. 

An even more precise distinction regarding the effects that make up the network's assortativity structure is possible if, in a third step, the supplementary measures $r^\omega_{(1,0)}$ and $r^\omega_{(0,1)}$ are included into the comparison. For example, since the connection effect is captured by the parameter $\alpha$, a comparison of $r^\omega_{(0,0)}$ with $r^\omega_{(1,0)}$, and $r^\omega_{(0,1)}$ with $r^\omega_{(1,1)}$, respectively, provides information on how the assortativity of the network varies with respect to the consideration of edge weights in terms of using weighted vertex values instead of unweighted ones (i.e. strengths rather than degrees). Particularly, if $r^\omega_{(0,0)} < r^\omega_{(1,0)}$, then incorporating edge weights leads to an increase in assortativity, suggesting that \textit{unweighted} connections are more assortative or less disassortative with respect to strengths as compared to degrees. Whereas, if $r^\omega_{(0,0)} > r^\omega_{(1,0)}$, then incorporating edge weights leads to a decrease in assortativity, suggesting that \textit{unweighted} connections are less assortative or more disassortative with respect to strengths as compared to degrees. Similarly, the connection effect can be interpreted for weighted connections, i.e., if $r^\omega_{(0,1)} < r^\omega_{(1,1)}$, then incorporating edge weights leads to an increase in assortativity, suggesting that \textit{weighted} connections are more assortative or less disassortative with respect to strengths as compared to degrees. As opposed to this, if $r^\omega_{(0,1)} > r^\omega_{(1,1)}$, then incorporating edge weights leads to a decrease in assortativity, suggesting that \textit{weighted} connections are less assortative or more disassortative with respect to strengths as compared to degrees.

By the same logic, since the amplification effect is captured by the patameter $\beta$, comparisons of $r^\omega_{(0,0)}$ with $r^\omega_{(0,1)}$, and $r^\omega_{(1,0)}$ with $r^\omega_{(1,1)}$, respectively, reveal how assortativity changes if edge weights are considered in terms of using weighted connections instead of unweighted ones in the computation of the assortativity coefficient for fixed vertex values (strengths or degrees). To be more precise, if $r^\omega_{(0,0)} < r^\omega_{(0,1)}$, then the increase in assortativity suggests that high weighted connections tend to be more assortative or less disassortative \textit{by degree} than low weighted ones. Vice versa, if $r^\omega_{(0,0)} > r^\omega_{(0,1)}$, there is a decrease in assortativity, suggesting that high weighted connections are less assortative or more disassortative \textit{by degree} than low weighted ones. Similarly, if $r^\omega_{(1,0)} > r^\omega_{(1,1)}$, then the increase in assortativity suggests that high weighted connections tend to be more assortative or less disassortative \textit{by strength} than low weighted ones. Again, if, vice versa, $r^\omega_{(1,0)} < r^\omega_{(1,1)}$, then the decrease in assortativity suggests that high weighted connections tend to be less assortative or more disassortative \textit{by strength} than low weighted ones.

\begin{mybox}[Procedure for Assessing a Network's Assortativity]{assessing_assortativity_box}
\begin{enumerate}
	\item Compute $r^\omega_{(\alpha, \beta)}$ for all four parameter combinations $(\alpha, \beta)$.

	\item An indication of the overall assortativity is given by the values of $r^\omega_{(0,0)} $ and $ r^\omega_{(1,1)}$:
		\begin{itemize}
			\item if values $ > 0$ $\Rightarrow$ assortative
			\item if values $ < 0$ $\Rightarrow$ disassortative
			\item if values $ = 0$ $\Rightarrow$ non-assortative
		\end{itemize}

	\item Obtain the edge weight effect on assortativity by comparing the generalised with the benchmark assortativity coefficient:
		\begin{itemize}
			\item if $r^\omega_{(0,0)} > r^\omega_{(1,1)}$ $\Rightarrow$ weighted network is less assortative
			\item if $r^\omega_{(0,0)} < r^\omega_{(1,1)}$ $\Rightarrow$ weighted network is more assortative
		\end{itemize}

	\item Interpret assortativity effects by comparing the generalised with the benchmark assortativity coefficient:
	\begin{itemize}
		\item \emph{Connection effect}
			\begin{itemize}
				\item if $ r^\omega_{(0,0)} (>) <  \, r^\omega_{(1,0)} $ $ \Rightarrow $
				most \textit{unweighted} connections tend to be (less) more assortative by strength as by degree
				\item if $ r^\omega_{(0,1)} (>) <  \, r^\omega_{(1,1)} $ $ \Rightarrow $
				most \textit{weighted} connections tend to be (less) more assortative by strength as by degree
			\end{itemize}

		\item \emph{Edge amplification effect}
			\begin{itemize}
				\item if $r^\omega_{(0,0)} (>) < \, r^\omega_{(0,1)}$ $\Rightarrow$ high weighted connections tend to be (less) more assortative \textit{by degree}
				\item if $r^\omega_{(1,0)} (>) < \, r^\omega_{(1,1)}$ $\Rightarrow$ high weighted connections tend to be (less) more assortative \textit{by strength}
			\end{itemize}
		\item Both effects
			\begin{itemize}
				\item if one or both effects operate in the same (different) direction with respect to the way in which edge weights are considered $\Rightarrow$ respective effect is consistent (inconsistent)
				\item if $r^\omega_{(1,1)} < \text{min}\big(r^\omega_{(1,0)},r^\omega_{(0,1)}\big) \vee r^\omega_{(1,1)} > \text{max}\big(r^\omega_{(1,0)},r^\omega_{(0,1)}\big)\Rightarrow$ consensual
				\item if $r^\omega_{(1,1)} \in \Big[\text{min}\big(r^\omega_{(1,0)},r^\omega_{(0,1)}\big),\text{max}\big(r^\omega_{(1,0)},r^\omega_{(0,1)}\big)\Big]\Rightarrow$ opposing
			\end{itemize}
	\end{itemize}
\end{enumerate}
\end{mybox}

If the assortativity of a network increases due to one of the two effects, we call the respective effect assortative, if, however, the assortativity of a network decreases, we call the respective effect disassortative. As can be seen from the above, both effects are twofold as they might operate differently with respect to the way in which edge weigths are considered. For example, for the same network, the assortativity might vary differently for \textit{unweighted} and \textit{weighted connections} if edge weights are considered via weighted vertex values as with the connection effect. The same holds true for the amplification effect where the assortativity with respect to both \textit{unweighted vertex values} as well as \textit{weighted vertex values} might vary differently if edge weights are considered in terms of weighted connections. We call the effects \textit{consistent} if, considered individually, they operate in the same way, and \textit{inconsistent} the other way round. For example, a connection effect which reduces the assortativity of the network for both unweighted as well as weighted connections if edge weights are considered with respect to weighted vertex values is considered \textit{consistent}, in this particular case consistently disassortative. Contrary to this, an amplification effect that increases assortativity by degree on the one hand but decreases assortativity by strength on the other is considered \textit{inconsistent}. In this case, there exists an assortative amplification effect with respect to unweighted vertex values but, at the same time, a disassortative amplification effect with respect to weighted vertex values.

Finally, we can determine whether the effects are \textit{consensual} or \textit{opposing}.
The effects are consensual if the assortativity coefficient for which one of the effects has already been taken into account increases or decreases even further if the other effect is additionally taken into account. This is the case if $r^\omega_{(1,1)} < \text{min}\big(r^\omega_{(1,0)},r^\omega_{(0,1)}\big)$ or $ r^\omega_{(1,1)} > \text{max}\big(r^\omega_{(1,0)},r^\omega_{(0,1)}\big)$.
If, however, $r^\omega_{(1,1)} \in \Big[\text{min}\big(r^\omega_{(1,0)},r^\omega_{(0,1)}\big),\text{max}\big(r^\omega_{(1,0)},r^\omega_{(0,1)}\big)\Big]$, then, this indicates that the effects are opposing because there is an effect that results in a more disassortative or assortative coefficient, respectively, if the other one is not considered, i.e., the impact of the first effect is reduced by the second. 

The outlined procedure is summarized in Box\,\ref{assessing_assortativity_box} and we will illustrate its application to empirical networks in Section \ref{section_application}.

\subsection{Assessing the Significance of Assortativity}\label{section_significance}

As mentioned above, there are four different ways of measuring assortativity in weighted networks. If we interpret each of them to be an estimator of the respective unknown population parameter, then computing the values for the coefficients based on a real network yields the corresponding point estimates. However, the associated estimation uncertainty is unknown, such that inference on the individual assortativity coefficients is infeasible, unless standard errors are computed, which is a challenging task in network analysis, as there is usually just one realisation of a real network and no sample of realisations available. Therefore, resampling methods such as the jackknife or bootstrap method are employed, which generate artificial samples of networks based on which an estimate of the standard error can be derived, cf. \cite{Quenouille, Tukey, Efron1979}. This allows to conduct significance tests for the respective assortativity coefficient.

Furthermore, when assessing a specific network characteristic, it is oftentimes of interest to decide whether the observed characteristic is due to some underlying social or organizational process or due to structural contraints, see \cite{Maslov2}. We will therefore compare the observed assortativity coefficients to the values we would have obtained if edges had formed randomly, i.e. the assortativity of a null model. The latter is obtained based on a link rewiring technique. 

In the following, we present the jackknife and bootstrap methods as well as the link rewiring technique adopted in this paper. Thereafter, we summarize our procedure for the statistical assessment of the assortativity coefficients. 

\subsubsection{Resampling Methods for Networks} \label{section_resampling}

The idea underlying the jackknife method is as follows: For a dataset that consists of $n$ sample variables, $n$ artificial subsamples are created by successively removing the $i$-th sample variable, $i = 1, \hdots, n$. For networks, there are different approaches of adopting the jackknife method. One can either consider the $n$ vertices of a network as sample variables and create subsamples by removing the vertices in turn, thus, a single subsample is the induced subgraph of the $(n - 1)$ remaining vertices, cf. \cite{Snijders}, or the jackknife method is applied to the $m$ edges, i.e., by removing the edges in turn as suggested by Newman \cite{Newman2003}.
For reasons of comparability, we will use the latter for our analysis, as it already has been adopted for the original assortativity coefficient, and since then, has also been used in order to estimate the standard errors of other network quantities such as \textit{reciprocity}, see \cite{Garlaschelli, Squartini}.
Thus, our jackknife estimate of the standard error of the generalised assortativity coefficient, denoted by $\hat{\sigma}_{r^\omega_{(\alpha, \beta)}, J}$, is computed as
\begin{align}
	\hat{\sigma}_{r^\omega_{(\alpha, \beta)}, J} = \sqrt{\sum_{i=1}^m \big(r^\omega_{(\alpha, \beta), (-i)}-r^\omega_{(\alpha, \beta)}\big)^2},\label{eq:jack_std_error}
\end{align}
where $ r^\omega_{(\alpha, \beta), (-i)} $ is the value of the assortativity coefficient for the network where the $i$-th edge has been removed. This allows to assess the uncertainty about the assortativity of a network by constructing corresponding confidence intervals. We compute a 95 percent jackknife confidence interval according to
\begin{align}
	CI_{0.95, J} = \big[r^\omega_{(\alpha, \beta)} - d, r^\omega_{(\alpha, \beta)} + d\big],\quad d = z_{0.975} \cdot \hat{\sigma}_{r^\omega_{(\alpha, \beta)}, J},
	\label{eq:jack_confidence_interval}
\end{align}
where $ z_{0.975} $ is the 97.5 percent quantile of the standard normal distribution.

Note that in large networks with many edges this approach can be computationally intensive.
In cases where computation times are prohibitively long it might be sensible to consider the jackknife with respect to the vertices instead, since the count of vertices is usually much lower than the count of edges.

Alternatively, the bootstrap method can be used, which, for large networks, outperforms the jackknife in terms of computational cost. 
The idea of the bootstrap is, again for a dataset that consists of $n$ sample variables, to consider the data as a population itself.
A subsample is then generated by sampling $n$ variables with replacement from the observed data. Thus, a subsample might contain multiple copies of some of the variables, and at the same time, no copies of some of the other variables.

In this paper, we follow the nonparametric approach of \,\cite{Snijders}, the so-called \textit{vertex bootstrap}
, and sample with replacement from the vertices of an observed network. 
More precisely, consider the weighted $n \times n$ adjacency matrix $\bm{W} = [w_{ij}]$ of the observed network, where $n$ is the number of vertices and the elements $w_{ij}$ represent the weights of the edges connecting vertices $i$ and $j$. If, however, $i$ and $j$ are not connected then $w_{ij} = 0$.\footnote{We focus on directed weighted networks. Nevertheless, the approach is capable of handling any kind of (un)directed and (un)weighted network.} A sample with replacement is drawn from the sequence of vertices $i = 1, \hdots, n$, and denoted by $i(1), \hdots, i(n)$.
A single bootstrap network is created by letting $\bm{W}^\ast = [w^\ast_{hk}]$, where its elements are obtained from the observed weighted adjacency matrix
\begin{align}
	w^\ast_{hk} = w_{i(h)i(k)},\quad i(h) \neq i(k).
\end{align}
In the case of $i(h) = i(k)$, i.e., $i(h)$ and $i(k)$ correspond to the same vertex in the observed network, the weight $w^\ast_{hk}$ is sampled randomly from the set of all observed edges, since self-edges or loops are usually not considered in real networks.
Thereafter, the generalised assortativity coefficient of the bootstrapped network $\hat{\theta}^\ast = r_{(\alpha, \beta)}^\omega(\bm{W}^\ast)$ is computed. Repeating the above procedure $B$ independent times yields an ensemble of $B$ bootstrap replications of the estimate of assortativity, $\hat{\theta}^\ast_1, \hdots, \hat{\theta}^\ast_B$, such that the bootstrap estimate of the standard error of the generalised assortativity coefficient can be obtained according to
\begin{align}
	\hat{\sigma}_{r^\omega_{(\alpha, \beta)}, B} = \hat{\sigma}_{\hat{\theta}, B} = \sqrt{\frac{1}{B-1}\sum_{b=1}^B \Big(\hat{\theta}^\ast_b- \bar{\hat{\theta}}^\ast\Big)^2},
	\label{eq:boot_std_error}
\end{align}
where $ \bar{\hat{\theta}}^\ast $ is the mean of the $B$ bootstrap replications of the assortativity coefficient. The corresponding 95 percent bootstrap confidence interval is given by\footnote{This is the standard bootstrap confidence interval, which is sometimes called the \textit{normal approximation} confidence interval, cf. \cite{Efron4,Davison}. It has the advantage that it can be compared to the jackknife confidence interval in Eq.\,(\ref{eq:jack_confidence_interval}). For our purpose, we verify its validity by analysing the histogram and Q-Q plots of the distribution of the bootstrap replications. The corresponding results are available from the authors upon request. Alternatively, statistical tests, such as the Jarque-Bera or Anderson-Darling, can be employed to check the normality assumption, though, in our case the bootstrap diagnostic plots were conclusive. If the normal distribution assumption is not appropriate, more advanced bootstrap confidence intervals can be used, cf. \cite{Efron2, Efron3, Efron4, DiCiccio}, and the excellent overview given in\,\cite{Davison}.}
\begin{align}
	CI_{0.95, B} = \big[r^\omega_{(\alpha, \beta)} - d, r^\omega_{(\alpha, \beta)} + d\big],\quad d = z_{0.975} \cdot \hat{\sigma}_{r^\omega_{(\alpha, \beta)}, B},
	\label{eq:boot_confidence_interval}
\end{align}

The computational cost of bootstrapping depends on the number of generated subsamples $B$. Indeed, the number of bootstrap samples $B$ has to be large enough to adequately approximate the distribution of the generalised assortativity coefficient, however, if $B$ is less than the number of vertices $n$ as well as the number of edges $m$ in a network, then the bootstrap requires less computation than both the jackknife with respect to vertices and the jackknife with respect to edges, cf. \cite{Cameron}.

For the sake of completeness, we report in our empirical analysis both, jackknife and bootstrap standard error estimates, together with the 95 percent confidence intervals for the generalised assortativity coefficient. We find that in almost all considered cases the results based on the bootstrap are in line with those based on the jackknife. In rare cases, where the results are ambiguous, we rely on the method that used a larger set of subsamples, i.e., we prefer the bootstrap over the jackknife in small networks $(B > m)$ and vice versa in large networks $(B < m)$.

\subsubsection{Generation of a Null Model by Link Rewiring}\label{subsubsection_null}

Newman \cite{Newman2003} gives an attempt at an explanation for the phenomenon of assortative mixing (by degree). A distinction is made between the degree correlations that originate from social or organizational processes (e.g. attraction or affiliation) and others that are artifacts resulting from structural constraints that are imposed on the type of network (e.g. structural disassortativity as discussed in\,\cite{Maslov2}).

To assess whether a network's assortativity is due to some underlying social or organizational process or due to structural contraints, we compare its observed value to the value we would have expected if edges had formed randomly, i.e. the assortativity of a null model, which we obtain based on a link rewiring technique. In doing so, we adopt the general approach of\,\cite{Maslov2} to detect and analyse topological patterns in networks, to the context of assortativity.\,\cite{Maslov2} suggest that a statistically significant deviation of a topological property of a network from the one of an appropriate null model presumably reflects that the property has real social or organizational origins.
Consequently, if there is no significant deviation, one will commit a mistake by attaching too much importance to the pattern as it appears to be a result from structural contraints, i.e., it appears to be random with respect to the type of network.

A null model is a random network that is matched for basic properties other than the one of interest, such as order, size and degree distribution, see \cite{Fornito}. As our focus is on weighted networks we expand these basic properties by the network's strength distribution and weight distribution. The null distribution is sampled by employing a switching based graph generating approach where Markov chains are used to generate an ensemble of randomized networks, see \cite{Ying,Milo}.

In order to create a single random network we apply the two step algorithm suggested by\,\cite{Rubinov} where, initially, the binary edges of the observed network are rewired such that the degree distribution is preserved. To this end we use the well-known algorithm by Maslov and Sneppen \cite{Maslov}. Afterwards, the original network's edge weights are assigned to the edges of the randomised network such that the observed strengths are closely approximated. This is done by randomly selecting an element $a_{uv}$ from the randomised network, with expected weight rank $i$ and assigning to it the $i$-th highest previously unassigned observed edge weight $w_{uv}$. The expected (unassigned) weight magnitude of an element $a_{uv}$ is $\tilde{e}_{uv} \propto \Big(s_u - \sum_h \tilde{w}_{uh}\Big)\Big(s_v - \sum_h \tilde{w}_{hv}\Big)$. Arranging $a_{uv}$ by $\tilde{e}_{uv}$ yields the expected weight rank $i$. After assigning $w_{uv}$ to $a_{uv}$ the pair is removed from further consideration. The remaining elements are then re-arranged by $\tilde{e}_{uv}$ and the procedure repeats by randomly selecting another element $a_{uv}$ until all elements have been assigned an observed edge weight. A random network generated by this algorithm preserves the degree sequence of the original network and, thus, the (in- and out-) degree distribution, exactly. It also preserves the weight distribution but not the weight sequence. Therefore, the observed strengths of the original network will only be closely approximated. However, a review of the relevant literature shows that, so far, there is no null model of weighted networks that preserves observed strengths exactly. We therefore follow\,\cite{Rubinov} and check whether the correlation between the pre- and post-randomization strength sequences is high. Moreover, the Kolmogorov-Smirnov two-sample test indicates that the pre- and post-randomization strength sequences follow the same distribution.\footnote{Results are available from the authors upon request.} We therefore conclude that the considered null model is appropriate for our purpose.

\subsubsection{Statistical Assessment of Assortativity}\label{subsubsection_statistical_assessment}

In the following we make use of the standard errors obtained from the jackknife and bootstrap method, in order to test for the significance of the generalised assortativity coefficient, and construct confidence intervals implied by the null model, in order to determine, whether the observed assortativity is due to organizational or social effects, or due to structural constraints. The procedure is summarized in Box\,\ref{assessing_significance_box}.

In particular, to test, whether the assortativity coefficient is significantly different from zero, we check whether the 95 percent jackknife or bootstrap confidence interval of the generalised assortativity coefficient, covers the value zero. If zero is not included, we conclude that the generalised assortativity coefficient is significantly different from zero at the 5 percent significance level. Furthermore, a comparison of the mean of the assortativity of the null model with the observed value of assortativity allows to assess the origins of the assortativity. More precisely, if the 95 percent confidence interval of the mean assortativity of the null model does not encompass the observed assortativity coefficient, we conclude that the assortativity is due to some social or organizational processes.
Vice versa, if the computed interval covers the observed assortativity coefficient, this indicates that the assortativity structure of the observed network is due to structural constraints, and thus random with respect to basic features of the network.

\begin{mybox}[Procedure for Assessing Significance of Assortativity]{assessing_significance_box}
	\begin{enumerate}
		\item Estimate $\hat{\sigma}_{r^\omega_{(\alpha, \beta)}}$ for all four parameter combinations $(\alpha, \beta)$ using a suitable method (e.g. jackknife method as in Eq.\,(\ref{eq:jack_std_error}) or bootstrap method as in Eq.\,(\ref{eq:boot_std_error}))
		\item Compute confidence intervals, $\text{CI}_{r^\omega_{(\alpha, \beta)}, S}$, for a predefined confidence level $S$, e.g. $S = 95\%$ (e.g. jackknife confidence intervals as in Eq.\,(\ref{eq:jack_confidence_interval}) or bootstrap confidence intervals as in Eq.\,(\ref{eq:boot_confidence_interval})), and interpret according to:
			\begin{itemize}
				\item if $0 \notin \text{CI}_{r^\omega_{(\alpha, \beta)}, S}$ $\Rightarrow$ statistically \textit{significant} assortative mixing
				\item if $0 \in \text{CI}_{r^\omega_{(\alpha, \beta)}, S}$ $\Rightarrow$ assortative mixing statistically \textit{insignificant}
			\end{itemize}
		\item Determine the distribution of the assortativity of a respective null model by a suitable method (e.g. link rewiring as described) and estimate its mean,  $r^\omega_{(\alpha, \beta), \text{rnd}}$
		\item Compute confidence intervals of the mean assortativity of the null model, $\text{CI}_{r^\omega_{(\alpha, \beta), \text{rnd}}, S}$, for a predefined confidence level $S$, e.g. $S = 95\%$, and interpret according to:
			\begin{itemize}
				\item if $r^\omega_{(\alpha, \beta)} \notin \text{CI}_{r^\omega_{(\alpha, \beta), \text{rnd}}, S}$ $\Rightarrow$ network's assortativity structure appears to have social or organizational origins
				\item if $r^\omega_{(\alpha, \beta)} \in \text{CI}_{r^\omega_{(\alpha, \beta), \text{rnd}}, S}$ $\Rightarrow$ network's assortativity structure appears to be random with respect to basic features of the network (e.g. size, order)
			\end{itemize}
	\end{enumerate}
\end{mybox}

\section{Application---Assortativity of Real-World Weighted Directed Networks}
\label{section_application}

In the following we apply our generalised assortativity coefficient to several (un)directed weighted real-world networks and illustrate its usefulness in assessing and interpreting the assortativity structure of these networks by incorporating weighted edges. To this end, we follow the procedure outlined in Box\,\ref{assessing_assortativity_box} and Box\,\ref{assessing_significance_box}. The analysed networks are taken from the website of the \textit{Koblenz Network Collection} project (KONECT), cf. \cite{KONECT}, and have also already been considered in previous literature. Tab.\,\ref{tab_assortativity_application} presents for each network the assortativity coefficients $r^\omega_{(\alpha, \beta)}$ for the different parameter combinations $(\alpha, \beta)$ along with the corresponding jackknife and bootstrap estimates of the standard error, $\hat{\sigma}_{J}$ and $\hat{\sigma}_{B}$, respectively, as well as the 95 percent confidence intervals, $\text{CI}_{0.95, J}$ and $\text{CI}_{0.95, B}$. Bootstrap results are based on $B = 1499$ boostrap replications.\footnote{
	Based on\,\cite{Davidson} where a minimum of $B = 399$ bootstrap replications for tests at the $0.05$ level and a minimum of $B = 1499$ bootstrap replications for tests at the $0.01$ level is suggested, we choose $B = 1499$, although we test at the $0.05$ level.
} We estimate the mean $r^\omega_\text{rnd}$ and standard deviation $\hat{\sigma}_{r^\omega_\text{rnd}}$ of the assortativity of the respective null models based on an ensemble of $1000$ randomisations of the observed network, where the number of switching steps per randomisation is set to $k = 20m$, as recommended in\,\cite{Fornito, Ying}.\footnote{Alternatively, $k = 100m$ can be chosen, cf. \cite{Milo}, but this is computationally more demanding.
} The lower and upper bounds of the 95 percent confidence intervals $\text{CI}_{r^\omega_\text{rnd}, 0.95}$ are obtained by computing the respective quantiles of the distribution of the randomised assortativity.
Subsequently, a detailed analysis of the assortativity structure of each of the networks is given.

The NetScience network is an undirected collaboration network of scientists working on network theory which has been constructed by Newman \cite{Newman2001}. A node in the network represents a scientist and an edge between two scientists indicates that both co-authored one or more publications. In total 2742 co-authorships of 1589 scientists have been included. The intensity of the relation between two scientists is incorporated by positive edge weights, which are defined as $ w_{ij} = \sum_k \nicefrac{\delta_i^k \delta_j^k}{(n_k - 1)} $, where $\delta_i^k = 1$ if scientist $i$ was co-author of paper $k$, and $n_k$ is the total number of co-authors of a paper $k$. As such, edge weights take into account that co-authors of large collaborations might know each other less than co-authors of smaller collaborations. Consequently, the interpretations of both vertex degree and vertex strength have to be considered carefully. In particular, vertex degree corresponds to the number of different co-authors scientist $i$ has collaborated with, whereas vertex strength corresponds to the number of papers scientist $i$ has co-authored with others, cf. \cite{Newman2001}.

\begin{center}
\singlespacing
  \begin{tiny}
    {\setlength{\tabcolsep}{2pt}
    \renewcommand{\arraystretch}{1.3}
    \begin{longtable}{
        llld{1.4}d{1.4}d{1.4}d{1.4}
      }
    \caption{Results for several weighted, directed and undirected real-world networks. Reported are: Generalised assortativity coefficient $r^\omega$, randomised assortativity coefficient $r^\omega_\text{rnd}$, expected jackknife and bootstrap standard errors on the observed assortativity, $\hat{\sigma}_{r^\omega,J}$ and $\hat{\sigma}_{r^\omega,B}$, respectively, expected standard error on the randomised assortativity $\hat{\sigma}_{r^\omega_\text{rnd}}$ as well as 95 percent jackknife and bootstrap confidence intervals, $\text{CI}_{r^\omega,0.95, J}$ and $\text{CI}_{r^\omega,0.95, B}$, for the observed assortativity coefficient, and 95 percent confidence intervals $\text{CI}_{r^\omega_\text{rnd}, 0.95}$ for the randomised assortativity coefficient, for all four parameter combinations $(\alpha, \beta)$. A detailed description of the networks is given in the text.}
      \label{tab_assortativity_application} \\
      \toprule
      \multirow{2}[2]{*}{Name} & \multirow{2}[2]{*}{Mode} &  \multirow{2}[2]{*}{Measure} & \multicolumn{2}{c}{$\alpha = 0$} & \multicolumn{2}{c}{$\alpha = 1$}\\ \cmidrule(lr){4-5}\cmidrule(lr){6-7}
      &&&
      \multicolumn{1}{c}{$\beta = 0$} & \multicolumn{1}{c}{$\beta = 1$} &
      \multicolumn{1}{c}{$\beta = 0$} & \multicolumn{1}{c}{$\beta = 1$} \\
      \midrule
    \endfirsthead
    \multicolumn{7}{c}%
    {\tablename\ \thetable\ -- \textit{Continued from previous page}} \\
    \toprule
    \multirow{2}[3]{*}{Name} & \multirow{2}[3]{*}{Mode} &  \multirow{2}[3]{*}{Measure} & \multicolumn{2}{c}{$\alpha = 0$} & \multicolumn{2}{c}{$\alpha = 1$}\\ \cmidrule(lr){4-5}\cmidrule(lr){6-7}
    &&&
    \multicolumn{1}{c}{$\beta = 0$} & \multicolumn{1}{c}{$\beta = 1$} &
    \multicolumn{1}{c}{$\beta = 0$} & \multicolumn{1}{c}{$\beta = 1$} \\
    \midrule
    \endhead
    \bottomrule \multicolumn{7}{r}{\textit{Continued on next page}} \\
    \endfoot
    \bottomrule
    \endlastfoot

	NetScience &&&& \\ \midrule
& undirected &&&& \\ \cmidrule{2-7}
&& $r^\omega$ & 0.4616 & 0.3405 & 0.1016 & 0.1928 \\
 && $r^\omega_\text{rnd}$ & -0.0436 & -0.0051 & -0.0691 & -0.0988 \\
 && $\hat{\sigma}_{r^\omega,J}$ & 0.0715 & 0.0618 & 0.0282 & 0.0527 \\
 && $\hat{\sigma}_{r^\omega,B}$ & 0.0944 & 0.1054 & 0.0978 & 0.1223 \\
 && $\hat{\sigma}_{r^\omega_\text{rnd}}$ & 0.0173 & 0.0339 & 0.0138 & 0.0305 \\
 && $\text{CI}_{r^\omega,0.95, J}$ & \multicolumn{1}{c}{$[ 0.3215 , 0.6017 ]$} &
 \multicolumn{1}{c}{$[ 0.2194 , 0.4616 ]$} &
 \multicolumn{1}{c}{$[ 0.0463 , 0.1569 ]$} &
 \multicolumn{1}{c}{$[ 0.0895 , 0.2961 ]$} \\
 && $\text{CI}_{r^\omega, 0.95, B}$ & \multicolumn{1}{c}{$[ 0.2766 , 0.6466 ]$} &
 \multicolumn{1}{c}{$[ 0.1339 , 0.5471 ]$} &
 \multicolumn{1}{c}{$[ -0.0901 , 0.2933 ]$} &
 \multicolumn{1}{c}{$[ -0.0469 , 0.4325 ]$} \\
 && $\text{CI}_{r^\omega_\text{rnd}, 0.95}$ & \multicolumn{1}{c}{$[ -0.0709 , -0.0143 ]$} &
 \multicolumn{1}{c}{$[ -0.0586 , 0.0498 ]$} &
 \multicolumn{1}{c}{$[ -0.0910 , -0.0462 ]$} &
 \multicolumn{1}{c}{$[ -0.1452 , -0.0459 ]$} \\
\pagebreak
Windsurfers &&&& \\ \midrule
& undirected &&&& \\ \cmidrule{2-7}
&& $r^\omega$ & -0.1470 & -0.0170 & -0.1710 & -0.0769 \\
 && $r^\omega_\text{rnd}$ & -0.2182 & -0.1187 & -0.1880 & -0.2266 \\
 && $\hat{\sigma}_{r^\omega,J}$ & 0.0654 & 0.1285 & 0.0481 & 0.1077 \\
 && $\hat{\sigma}_{r^\omega,B}$ & 0.0465 & 0.0833 & 0.0343 & 0.0901 \\
 && $\hat{\sigma}_{r^\omega_\text{rnd}}$ & 0.0324 & 0.0740 & 0.0251 & 0.0760 \\
 && $\text{CI}_{r^\omega,0.95, J}$ & \multicolumn{1}{c}{$[ -0.2752 , -0.0188 ]$} &
 \multicolumn{1}{c}{$[ -0.2689 , 0.2349 ]$} &
 \multicolumn{1}{c}{$[ -0.2653 , -0.0767 ]$} &
 \multicolumn{1}{c}{$[ -0.2880 , 0.1342 ]$} \\
 && $\text{CI}_{r^\omega, 0.95, B}$ & \multicolumn{1}{c}{$[ -0.2381 , -0.0559 ]$} &
 \multicolumn{1}{c}{$[ -0.1803 , 0.1463 ]$} &
 \multicolumn{1}{c}{$[ -0.2382 , -0.1038 ]$} &
 \multicolumn{1}{c}{$[ -0.2535 , 0.0997 ]$} \\
 && $\text{CI}_{r^\omega_\text{rnd}, 0.95}$ & \multicolumn{1}{c}{$[ -0.2711 , -0.1654 ]$} &
 \multicolumn{1}{c}{$[ -0.2434 , -0.0016 ]$} &
 \multicolumn{1}{c}{$[ -0.2299 , -0.1492 ]$} &
 \multicolumn{1}{c}{$[ -0.3530 , -0.0965 ]$} \\

Macaques &&&& \\ \midrule
& out-in &&&& \\ \cmidrule{2-7}
&& $r^\omega$ & -0.3709 & -0.3801 & -0.2479 & -0.2578 \\
 && $r^\omega_\text{rnd}$ & -0.1475 & -0.0394 & -0.1525 & -0.0526 \\
 && $\hat{\sigma}_{r^\omega,J}$ & 0.0377 & 0.0483 & 0.0381 & 0.0476 \\
 && $\hat{\sigma}_{r^\omega,B}$ & 0.0555 & 0.0638 & 0.0533 & 0.0668 \\
 && $\hat{\sigma}_{r^\omega_\text{rnd}}$ & 0.0174 & 0.0240 & 0.0173 & 0.0242 \\
 && $\text{CI}_{r^\omega,0.95, J}$ & \multicolumn{1}{c}{$[ -0.4448 , -0.2970 ]$} &
 \multicolumn{1}{c}{$[ -0.4748 , -0.2854 ]$} &
 \multicolumn{1}{c}{$[ -0.3226 , -0.1732 ]$} &
 \multicolumn{1}{c}{$[ -0.3511 , -0.1645 ]$} \\
 && $\text{CI}_{r^\omega, 0.95, B}$ & \multicolumn{1}{c}{$[ -0.4797 , -0.2621 ]$} &
 \multicolumn{1}{c}{$[ -0.5051 , -0.2551 ]$} &
 \multicolumn{1}{c}{$[ -0.3524 , -0.1434 ]$} &
 \multicolumn{1}{c}{$[ -0.3887 , -0.1269 ]$} \\
 && $\text{CI}_{r^\omega_\text{rnd}, 0.95}$ & \multicolumn{1}{c}{$[ -0.1765 , -0.1201 ]$} &
 \multicolumn{1}{c}{$[ -0.0790 , -0.0010 ]$} &
 \multicolumn{1}{c}{$[ -0.1809 , -0.1238 ]$} &
 \multicolumn{1}{c}{$[ -0.0922 , -0.0128 ]$} \\

& out-out &&&& \\ \cmidrule{2-7}
&& $r^\omega$ & 0.4162 & 0.4294 & 0.3145 & 0.3251 \\
 && $r^\omega_\text{rnd}$ & 0.1022 & 0.0375 & 0.0877 & 0.0269 \\
 && $\hat{\sigma}_{r^\omega,J}$ & 0.0400 & 0.0491 & 0.0424 & 0.0513 \\
 && $\hat{\sigma}_{r^\omega,B}$ & 0.0551 & 0.0641 & 0.0507 & 0.0559 \\
 && $\hat{\sigma}_{r^\omega_\text{rnd}}$ & 0.0200 & 0.0259 & 0.0195 & 0.0246 \\
 && $\text{CI}_{r^\omega,0.95, J}$ & \multicolumn{1}{c}{$[ 0.3378 , 0.4946 ]$} &
 \multicolumn{1}{c}{$[ 0.3332 , 0.5256 ]$} &
 \multicolumn{1}{c}{$[ 0.2314 , 0.3976 ]$} &
 \multicolumn{1}{c}{$[ 0.2246 , 0.4256 ]$} \\
 && $\text{CI}_{r^\omega, 0.95, B}$ & \multicolumn{1}{c}{$[ 0.3082 , 0.5242 ]$} &
 \multicolumn{1}{c}{$[ 0.3038 , 0.5550 ]$} &
 \multicolumn{1}{c}{$[ 0.2151 , 0.4139 ]$} &
 \multicolumn{1}{c}{$[ 0.2155 , 0.4347 ]$} \\
 && $\text{CI}_{r^\omega_\text{rnd}, 0.95}$ & \multicolumn{1}{c}{$[ 0.0698 , 0.1354 ]$} &
 \multicolumn{1}{c}{$[ -0.0055 , 0.0810 ]$} &
 \multicolumn{1}{c}{$[ 0.0564 , 0.1197 ]$} &
 \multicolumn{1}{c}{$[ -0.0127 , 0.0678 ]$} \\

& in-in &&&& \\ \cmidrule{2-7}
&& $r^\omega$ & 0.4030 & 0.4586 & 0.2277 & 0.2828 \\
 && $r^\omega_\text{rnd}$ & 0.1032 & 0.0260 & 0.0771 & 0.0132 \\
 && $\hat{\sigma}_{r^\omega,J}$ & 0.0386 & 0.0440 & 0.0468 & 0.0520 \\
 && $\hat{\sigma}_{r^\omega,B}$ & 0.0566 & 0.0677 & 0.0471 & 0.0570 \\
 && $\hat{\sigma}_{r^\omega_\text{rnd}}$ & 0.0194 & 0.0246 & 0.0198 & 0.0263 \\
 && $\text{CI}_{r^\omega,0.95, J}$ & \multicolumn{1}{c}{$[ 0.3273 , 0.4787 ]$} &
 \multicolumn{1}{c}{$[ 0.3724 , 0.5448 ]$} &
 \multicolumn{1}{c}{$[ 0.1360 , 0.3194 ]$} &
 \multicolumn{1}{c}{$[ 0.1809 , 0.3847 ]$} \\
 && $\text{CI}_{r^\omega, 0.95, B}$ & \multicolumn{1}{c}{$[ 0.2921 , 0.5139 ]$} &
 \multicolumn{1}{c}{$[ 0.3259 , 0.5913 ]$} &
 \multicolumn{1}{c}{$[ 0.1354 , 0.3200 ]$} &
 \multicolumn{1}{c}{$[ 0.1711 , 0.3945 ]$} \\
 && $\text{CI}_{r^\omega_\text{rnd}, 0.95}$ & \multicolumn{1}{c}{$[ 0.0723 , 0.1360 ]$} &
 \multicolumn{1}{c}{$[ -0.0144 , 0.0679 ]$} &
 \multicolumn{1}{c}{$[ 0.0445 , 0.1100 ]$} &
 \multicolumn{1}{c}{$[ -0.0301 , 0.0559 ]$} \\

& in-out &&&& \\ \cmidrule{2-7}
&& $r^\omega$ & -0.4884 & -0.5214 & -0.3933 & -0.4195 \\
 && $r^\omega_\text{rnd}$ & -0.0745 & -0.0320 & -0.0586 & -0.0287 \\
 && $\hat{\sigma}_{r^\omega,J}$ & 0.0234 & 0.0300 & 0.0283 & 0.0391 \\
 && $\hat{\sigma}_{r^\omega,B}$ & 0.0505 & 0.0586 & 0.0440 & 0.0545 \\
 && $\hat{\sigma}_{r^\omega_\text{rnd}}$ & 0.0211 & 0.0281 & 0.0214 & 0.0283 \\
 && $\text{CI}_{r^\omega,0.95, J}$ & \multicolumn{1}{c}{$[ -0.5343 , -0.4425 ]$} &
 \multicolumn{1}{c}{$[ -0.5802 , -0.4626 ]$} &
 \multicolumn{1}{c}{$[ -0.4488 , -0.3378 ]$} &
 \multicolumn{1}{c}{$[ -0.4961 , -0.3429 ]$} \\
 && $\text{CI}_{r^\omega, 0.95, B}$ & \multicolumn{1}{c}{$[ -0.5874 , -0.3894 ]$} &
 \multicolumn{1}{c}{$[ -0.6363 , -0.4065 ]$} &
 \multicolumn{1}{c}{$[ -0.4795 , -0.3071 ]$} &
 \multicolumn{1}{c}{$[ -0.5263 , -0.3127 ]$} \\
 && $\text{CI}_{r^\omega_\text{rnd}, 0.95}$ & \multicolumn{1}{c}{$[ -0.1084 , -0.0405 ]$} &
 \multicolumn{1}{c}{$[ -0.0793 , 0.0145 ]$} &
 \multicolumn{1}{c}{$[ -0.0919 , -0.0243 ]$} &
 \multicolumn{1}{c}{$[ -0.0748 , 0.0179 ]$} \\

\end{longtable}
    }
\end{tiny}
\end{center}

According to our empirical results, the NetScience network is an overall assortative network indicating that scientists have a tendency to collaborate with others that are similar based on the number of co-authors (degree) or based on the number of papers they have been coauthors of (strength), since both the benchmark assortivity coefficient, $r^\omega_{(0,0)} = 0.4616$, as well as the generalised assortativity coefficient, $r^\omega_{(1,1)} = 0.1928$, have positive values. However, since $r^\omega_{(0,0)} > r^\omega_{(1,1)} $, the network is less assortative if edge weigths are considered. On the one hand, this is partly due to a consistently disassortative connection effect, as $r^\omega_{(0,0)} = 0.4616 > r^\omega_{(1,0)} = 0.1015$ as well as $r^\omega_{(0,1)} = 0.3407 > r^\omega_{(1,1)} = 0.1928 $, which indicates that most interconnected scientists tend to be more similar based on the number of co-authors they have collaborated with and less similar regarding the number of co-authored papers. On the other hand, there is a disassortative amplification effect when degrees are used as vertex values, since $r^\omega_{(0,0)} = 0.4616 > r^\omega_{(0,1)} = 0.3407$, suggesting that the stronger co-author relationships persist between scientists that are less similar based on the number of co-authors they have collaborated with. However, since $ r^\omega_{(1,0)} = 0.1015 < r^\omega_{(1,1)} = 0.1928 $, the amplification effect is inconsistent and, thus, there is an assortative amplification effect when strengths are used as vertex values, which means that the stronger co-author relationships tend to persist between scientists that are more similar based on the number of papers they have published with others. Apparently, both effects are opposing as $r^\omega_{(1,1)} = 0.1928 \in \big[\text{min}\big(r^\omega_{(1,0)},r^\omega_{(0,1)}\big) = 0.1015 ,\text{max}\big(r^\omega_{(1,0)},r^\omega_{(0,1)}\big) = 0.3405\big]$. Consequently, scientists tend to collaborate with others that are either different based on the number of co-authors they have collaborated with or that are similar based on the number of papers they have published.

In order to assess the significance of the above results, we compute 95 percent jackknife and bootstrap confidence intervals for the coefficients and find that the NetScience network is significantly assortative. Moreover, as the 95 percent confidence intervals of the randomised assortativity do not cover the respective observed assortativity, i.e., $r^\omega_{(0,0)} = 0.4616 \notin \text{CI}_{r^\omega_{(0,0), \text{rnd}}, 0.95} = [ -0.0709 , -0.0143 ]$ and $r^\omega_{(1,1)} = 0.1928 \notin \text{CI}_{r^\omega_{(1,1), \text{rnd}}, 0.95} = [ -0.1452 , -0.0459 ]$, we can conclude at the 5 percent significance level, that the observed strong assortative structure is due to some social or sociological process. The network would have been disassortative if edges had formed randomly.\footnote{For the measures $r^\omega_{(1,0)}$ and $r^\omega_{(1,1)}$ the jackknife and the bootstrap intervals seem inconclusive. The respective coefficients differ indeed significantly from the null assortativity, though the bootstrap intervals of the observed assortativity encompass zero. However, in these cases we rely on the jackknife intervals as, for the NetScience network, they are based on the larger set of subsamples, as $B = 1499 < m = 2742$.}

The Windsurfers network is an undirected network which is formed by data that was collected while studying the social behaviour of 43 windsurfers on a beach in southern California during the fall of 1986, see\, \cite{Freeman}. A node in the network represents a windsurfer and an edge between two windsurfers indicates interpersonal contact.
Information on the frequency of this interpersonal contact is incorporated by positive edge weigths where a high edge weight indicates a more frequent contact and vice versa.
An edge weight of $1$ indicates a onetime contact. Thus, the degree of a vertex corresponds to the number of acquaintanceships, whereas vertex strength corresponds to the frequency of encounters.
The network consists of 336 weighted edges.

The Windsurfers network is overall disassortative, suggesting that there is a tendency that windsurfers connect to other windsurfers that are more interconnected than themselves. However, the weighted network is less disassortative, as can be seen by comparing the benchmark assortativity coefficient with the generalised assortativity coefficient, $r^\omega_{(0,0)} = -0.1470 < r^\omega_{(1,1)} = -0.0769$. Analysing both effects, the network indeed shows a consistently disassortative connection effect, which indicates that most interconnected windsurfers differ by the number of acquaintanceships, but differ even more by the frequency of encounters, since $ r^\omega_{(0,0)} = -0.1470 > r^\omega_{(1,0)} = -0.1710 $ and $ r^\omega_{(0,1)} = -0.0170 > r^\omega_{(1,1)} = -0.0769 $ and, thus, most connections persist between two windsurfers where one is more interconnected than the other. However, there is a consistently assortative edge amplification effect, since $ r^\omega_{(0,0)} = -0.1470 < r^\omega_{(0,1)} = -0.0170 $ and $ r^\omega_{(1, 0)} = -0.1710 < r^\omega_{(1,1)} = -0.0769 $, indicating that, although the network is overall disassortative, the high weighted connections tend to be rather assortative, i.e., windsurfers tend to stay in touch with others more frequently, mostly, if they are as interconnected as themselves, for example, with other windsurfers that either have an equal number of acquaintances or have equally frequent encounters. Since $r^\omega_{(1,1)} = -0.0769 \in \big[\text{min}\big(r^\omega_{(1,0)},r^\omega_{(0,1)}\big) = -0.1710 ,\text{max}\big(r^\omega_{(1,0)},r^\omega_{(0,1)}\big) = -0.0170\big]$, we have opposing effects for the Windsurfers network. Additionally, because both effects are consistent and because the magnitude of disassortativity is reduced by incorporating both effects, we can reason that the edge amplification effect might be the stronger one.

Considering the significance of the assortativity, the findings are somewhat inconclusive. On the one hand, the respective 95 percent confidence intervals of the randomised assortativity,  $\text{CI}_{r^\omega_{(\alpha,\beta), \text{rnd}}, 0.95}$, do not cover the values of $r^\omega_{(0,0)}$ and $r^\omega_{(1,1)}$, on the other hand, they cover the values of $r^\omega_{(0,1)}$ and $r^\omega_{(1,0)}$ indicating that the windsurfers network is indeed significantly more disassortative than we would expect if edges had formed randomly, but our previous conclusions regarding the connection effect as well as the amplification effect have to be questioned, as the auxiliary measures $r^\omega_{(0,1)}$ and $r^\omega_{(1,0)}$ appear to be insignificant with respect to the null model. Additionally, both the jackknife as well as the bootstrap 95 percent confidence intervals of the observed assortativity $r^\omega$ overlap the respective null assortivity coefficient $r^\omega_\text{rnd}$, for all parameter combinations $(\alpha, \beta)$ indicating that the disassortative structure of the Windsurfers network might also be, at least partially, structural and dependent on the type of the network.

The Macaques network is a directed network which is formed by data that was collected while studying the dominance relationships of 62 female Japanese monkeys (Macaca fuscata) during the nonmating season from April to early October 1976, see\, \cite{Takahata}.

A node in the network represents a specific monkey and directed edges between the nodes represent dominance relationships. Thus, an edge connecting two monkeys points from the dominating monkey to the one which has been dominated during an encounter where food was involved, and edge weights indicate how often such encounters happened. Since the network is directed we can differentiate between in- and out-degrees as well as in- and out-strengths. Thus, four different modes of assortativity are possible, which are denoted by \textit{out-in, out-out, in-in} and \textit{in-out}, where \textit{out-in}, for example, corresponds to the correlation between the excess out- and in-degrees (or strenghts) of two interconnected monkeys. The out-degree of a monkey corresponds to the number of different other monkeys it has dominated during an encounter, whereas its in-degree corresponds to the number of different other monkeys it has been dominated by. The out-strength, on the other hand, corresponds to the number of times where a monkey has dominated others, and the in-strength corresponds to the number of times it has been dominated by others. In total, the network consists of 1187 weighted edges. The edge weight sum of 2435 corresponds to the total number of observed encounters.

The network is overall \textit{out-in} as well as \textit{in-out} disassortative, since all of the measures $ r^\omega_{(0,0), \text{out-in}} = -0.3709 $, $ r^\omega_{(1,1), \text{out-in}} = -0.2578 $, $ r^\omega_{(0,0), \text{in-out}} = -0.4884 $ as well as $ r^\omega_{(1,1), \text{in-out}} = -0.4195 $ are negative.
This implies that monkeys who dominate many others, or dominate others more frequently, preferably dominate other monkeys who are dominated by only a few, or are dominated less frequently.
Vice versa, monkeys who dominate few others, or dominate others less frequently, tend to dominate other monkeys who are dominated by many others, or are dominated more frequently.
Comparing the respective benchmark with the generalised assortativity coefficient reveals that a full consideration of edge weights reduces the overall disassortativity of the network for both modes of assortativity, although not as much for the \textit{in-out} mode, as the difference between $r^\omega_{(0,0), \text{in-out}}$ and $r^\omega_{(1,1), \text{in-out}}$ compared to the difference between $r^\omega_{(0,0), \text{out-in}}$ and $r^\omega_{(1,1), \text{out-in}}$ shows.
At the same time, the network exhibits assortative tendencies with respect to the \textit{out-out} and \textit{in-in} modes, since $ r^\omega_{(0,0), \text{out-out}} = 0.4162 $, $ r^\omega_{(1,1), \text{out-out}} = 0.3251 $, $ r^\omega_{(0,0), \text{in-in}} = 0.4030 $ as well as $ r^\omega_{(1,1), \text{in-in}} = 0.2828 $ are all positive.
This indicates that dominating monkeys tend to dominate other dominating monkeys. Also inferior monkeys usually dominate other inferior monkeys.
The network is less \textit{out-out} as well as \textit{in-in} assortative if edge weights are considered, since $r^\omega_{(0,0), \text{out-out}} > r^\omega_{(1,1), \text{out-out}}$ and $r^\omega_{(0,0), \text{in-in}} > r^\omega_{(1,1), \text{in-in}}$.
The decrease in disassortativity for the \textit{out-in} and \textit{in-out} modes as well as the decrease in assortativity for the \textit{out-out} and \textit{in-in} modes can be explained by analysing the respective connection and amplification effects.
There is a consistently assortative connection effect for the \textit{out-in} and \textit{in-out} modes, since $r^\omega_{(0,0), \text{out-in}} = -0.3709 < r^\omega_{(1,0), \text{out-in}} = -0.2479$ and $r^\omega_{(0,1), \text{out-in}} = -0.3801 < r^\omega_{(1,1), \text{out-in}} = -0.2578$ as well as $r^\omega_{(0,0), \text{in-out}} = -0.4884 < r^\omega_{(1,0), \text{in-out}} = -0.3933$ and $r^\omega_{(0,1), \text{in-out}} = -0.5214 < r^\omega_{(1,1), \text{in-out}} = -0.4195$, but a consistently disassortative connection effect for the \textit{out-out} and \textit{in-in} modes, as $r^\omega_{(0,0), \text{out-out}} = 0.4162 > r^\omega_{(1,0), \text{out-out}} = 0.3145$ and $r^\omega_{(0,1), \text{out-out}} = 0.4294 > r^\omega_{(1,1), \text{out-out}} = 0.3251$ as well as $r^\omega_{(0,0), \text{in-in}} = 0.4030 > r^\omega_{(1,0), \text{in-in}} = 0.2277$ and $r^\omega_{(0,1), \text{in-in}} = 0.4586 > r^\omega_{(1,1), \text{in-in}} = 0.2828$.
This means that, for the \textit{out-in} and \textit{in-out} modes, \textit{most} interconnected monkeys differ less based on the number of times they have dominated or have been dominated by others, and vice versa, respectively, compared to the number of different monkeys they dominated or have been dominated by.
On the contrary, for the \textit{out-out} and \textit{in-in} modes, \textit{most} interconnected monkeys differ more based on both the number of times they have dominated others as well as the number of times they have been dominated by others compared to the number of different monkeys they dominated or have been dominated by, respectively.
Furthermore, there is a consistently disassortative amplification effect for the \textit{out-in} and the \textit{in-out} modes, because $r^\omega_{(0,0), \text{out-in}} = -0.3709 > r^\omega_{(0,1), \text{out-in}} = -0.3801$ and $r^\omega_{(1,0), \text{out-in}} = -0.2479 > r^\omega_{(1,1), \text{out-in}} = -0.2578$ as well as $r^\omega_{(0,0), \text{in-out}} = -0.4884 > r^\omega_{(0,1), \text{in-out}} = -0.5214$ and $r^\omega_{(1,0), \text{in-out}} = -0.3933 > r^\omega_{(1,1), \text{in-out}} = -0.4195$, but a consistently assortative amplification effect for the \textit{out-out} and \textit{in-in} modes, since $r^\omega_{(0,0), \text{out-out}} = 0.4162 < r^\omega_{(0,1), \text{out-out}} = 0.4294$ and $r^\omega_{(1,0), \text{out-out}} = 0.3145 < r^\omega_{(1,1), \text{out-out}} = 0.3251$ as well as $r^\omega_{(0,0), \text{in-in}} = 0.4030 < r^\omega_{(0,1), \text{in-in}} = 0.4586$ and $r^\omega_{(1,0), \text{in-in}} = 0.2277 < r^\omega_{(1,1), \text{in-in}} = 0.2828$. Indicating that dominant monkeys \textit{less frequently} dominate inferior monkeys, and vice versa. Also, monkeys dominate others that are similarly dominant or inferior, respectively, \textit{more frequently}.
The connection effect and the amplification effect are opposing for all four modes, as can be seen by the fact that if one is assortative the other is disassortative and vice versa, as well as by the fact that $r^\omega_{(1,1)} \in \big[\text{min}\big(r^\omega_{(1,0)},r^\omega_{(0,1)}\big),\text{max}\big(r^\omega_{(1,0)},r^\omega_{(0,1)}\big)\big]$ for all four modes. Additionally, since all effects are consistent, and since the weighted network is less disassortative for the \textit{out-in} and \textit{in-out} modes, but less assortative for the \textit{out-out} and \textit{in-in} modes, the connection effect can be regarded as the stronger effect for all four modes.
Finally, since all coefficients are significantly different from zero, and the 95 percent confidence intervals of the randomised assortativity do not cover any observed coefficient, we can conclude that the observed assortativity structure of the Macaques network has real social or sociological or organizational origins rather than being random with respect to basic network characteristics such as order, size, distribution of degrees, strengths or weights.

Based on the preceding analysis of the assortativity of the weighted example real-world networks, precise statements of their network topology are now possible. Using the example of the Macaques ---without having actually plotted the network graphically in advance, but having verified our observations afterwards--- we conclude that the network has a multi-tiered, almost tree-like hierarchical structure, that branches out into star-like configurations, where cycles are possible for the lower tiers, and where the higher weighted edges tend to form either between vertices where the out-degree or -strength of the one is different from the in-degree or -strength of the other, and vice versa, or between vertices of similar in-degree or -strength as well as vertices of similar out-degree or -strength. Admittedly, these observations also could have been obtained by studying the (weighted) adjacency matrix of the network, moreover, they are quite unsurprising even, given that dominance relationships between animals have been exhaustively studied for several species in the past. However, the fact that our proposed procedure of assessing the assortativity of a weighted network reveals the structural features of each example network precisely is exactly what we aim for. We think that in the case of networks which are larger by several orders of magnitude than the ones we considered for illustrative purposes in this paper, and for which the topology is not-well known, our generalised assortativity coefficient will provide useful insights.

\section{Conclusions and Future Work}

In this paper we have shown that assortativity, the tendency of vertices to bond with others based on similarities (usually excess vertex degree), in weighted networks is more complex than in unweighted networks.
Previously published research focuses on seeking a single measure that describes the underlying assortativity structure.
We pointed out, however, that focusing on a single measure might lead to information loss, and, therefore, proposed a generalised assortativity coefficient that nests previous measures and that utilises available information at the best.
To this end, we proposed to use as vertex values excess vertex strength, which has never been considered in the assortativity literature so far and which is the generalisation of excess vertex degree in weighted networks. We broke down assortativity in weighted networks into its components and identified two mechanisms that essentially affect the assortativity structure of a network, which we refer to as the connection effect as well as the amplification effect. Furthermore, we provided procedures that allow for a detailed interpretation and assessment of assortativity in weighted networks as well as for the assessments of its statistical significance. For the latter we introduced appropriate resampling and link rewiring techniques.
We demonstrated the application and usefulness of our generalised assortativity coefficient for assessing and interpreting the assortativity of three commonly used weighted real-world networks, both directed and undirected.

Based on our developments in this paper, it will be interesting to extend the concept of generalised assortativity to relative assortativity. This would allow to compare the values of the assortativity coefficients across different networks and would be along the lines of\,\cite{Squartini}, who have introduced a relative measure for the reciprocity of a network. Moreover, we plan to extend the concept of local assortativity, see\,\cite{Piraveenan}, according to our definition of generalised assortativity for weighted networks. We will address these points in future research.

\appendix

\section*{Appendix}

\subsection*{Excess (Out- or In-) Strengths in Directed Weighted Networks -- An Example}

\begin{minipage}{\linewidth}
	\begin{minipage}[b]{0.49\linewidth}
		\centering
		\begin{tikzpicture}[scale=1.75]
		\Vertex[label=A, color=lightgray]{A}
		\Vertex[label=D, x=0, y=1,color=lightgray]{D}
		\Vertex[label=E, x=0, y=-1,color=lightgray]{E}
		\Vertex[label=B, x=1,color=lightgray]{B}
		\Vertex[label=F, x=1,y=1,color=lightgray]{F}
		\Vertex[label=G, x=1,y=-1,color=lightgray]{G}
		\Vertex[label=H, x=2,y=0,color=lightgray]{H}
		\Vertex[label=C, x=-1,y=0,color=lightgray]{C}

		\Edge[lw=1pt, label=10, Direct](A)(B)
		\Edge[lw=1pt, label=3, Direct](A)(E)
		\Edge[lw=1pt, label=1, Direct](C)(A)
		\Edge[lw=1pt, label=2, Direct](D)(A)
		\Edge[lw=1pt, label=4, Direct](B)(F)
		\Edge[lw=1pt, label=5, Direct](B)(G)
		\Edge[lw=1pt, label=6, Direct](H)(B)

		\end{tikzpicture}
		
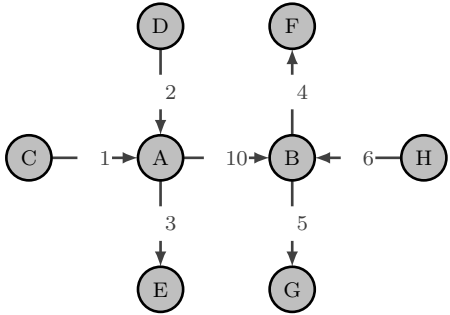
\captionof{figure}{A sample weighted directed network, adapted from \cite{Yuan}.\\}
		\label{fig:sample_weighted_network}
	\end{minipage}
	\hfill
	\begin{minipage}[b]{0.49\linewidth}
		\centering
		\setlength{\tabcolsep}{3pt}
		\begin{tabular}{cccccccccccc}
		\toprule
		$i$ & $l_i$ & $m_i$ & $\omega_i$ & & $\tilde{s}^{\textit{out}}_{(l,i)}$ & $\tilde{s}^{\textit{in}}_{(m,i)}$ & & $s^{\textit{out}}_{l}$ &  $s^{\textit{in}}_{l}$ & $s^{\textit{in}}_{m}$ & $s^{\textit{out}}_{m}$\\
		\cmidrule{1-4}\cmidrule{6-7}\cmidrule{9-12}
		1 & A & B & 10 && 3 & 6 && 13 & 3 & 16 & 9 \\
		2 & H & B & 6 && 0 & 10 && 6 & 3 & 16 & 0 \\
		3 & B & G & 5 && 4 & 0 && 9 & 16 & 5 & 0 \\
		4 & B & F & 4 && 5 & 0 && 9 & 16 & 4 & 0 \\
		5 & A & E & 3 && 10 & 0 && 13 & 0 & 3 & 13 \\
		6 & D & A & 2 && 0 & 1 && 2 & 0 & 3 & 13 \\
		7 & C & A & 1 && 0 & 2 && 1 & 0 & 3 & 9 \\
		\bottomrule
		\end{tabular}
		\captionof{table}{Edge list of network from Fig.\,\ref{fig:sample_weighted_network} expanded by excess and total out- and in-strengths.}
		\label{tab:sample_weighted_network}
	\end{minipage}
\end{minipage}%

\vspace{.5cm}
In the following we illustrate the consequences of using total strengths rather than excess strengths for assessing a network's assortativity structure. To this end we consider as an example the weighted and directed network considered in\,\cite{Yuan}, which is depicted in Fig.\,\ref{fig:sample_weighted_network}.
The directed edges are marked by their weights.
For example, the first edge of the network connects vertices A and B with an edge of weight $10$ pointing from A to B, i.e., $\omega_1 = w_{AB} = 10$.
The seventh edge of the network points from C to A with an edge weight of 1, i.e., $\omega_7 = w_{CA} = 1$.

The corresponding edge list of the network is depicted in Tab.\,\ref{tab:sample_weighted_network}. The edge list is expanded by the excess out- and in-strenghts as well as the total out- and in-strengths.
The excess out-strength of the source end $l$ of edge $i$ is defined as its total out-strength less the edge weight, i.e., $\tilde{s}^{\textit{out}}_{(l,i)} = s^{\textit{out}}_l - \omega_i$.
Similarly, the excess in-strength of the target end $m$ of edge $i$ is defined as its total in-strength less the edge weight, i.e., $\tilde{s}^{\textit{in}}_{(m,i)} = s^{\textit{in}}_m - \omega_i$.
The excess in-degree of the source end $l$ of edge $i$ as well as the excess out-strength of the target end $m$ of edge $i$ are not seperately listed in Tab.\,\ref{tab:sample_weighted_network}, since $\tilde{s}^{\textit{in}}_{(l,i)} = s^{\textit{in}}_l$ and $\tilde{s}^{\textit{out}}_{(m,i)} = s^{\textit{out}}_m$, respectively, as reasoned in Section\,\ref{section_definition}.
The total strengths are obtained as usual.
For example, the excess out-strength of vertex A with respect to the first edge is given by $\tilde{s}^{\textit{out}}_{(A,1)} = 13 - 10 = 3$ and, at the same time, the excess out-strength of vertex A with respect to the fifth edge is given by $\tilde{s}^{\textit{out}}_{(A,5)} = 13 - 3 = 10$.
Also, the excess in-strength of vertex A with respect to the seventh edge is given by $\tilde{s}^{\textit{in}}_{(A,7)} = 3 - 1 = 2$ and, at the same time, the excess in-strength of vertex A with respect to the sixth edge is $\tilde{s}^{\textit{in}}_{(A,6)} = 3 - 2 = 1$. This clearly shows that the excess (out- or in-) strengths of the ends of an edge depend on the weight of the edge that is currently considered. As opposed to this, the total strengths of the ends remain the same for all edges.

We will now illustrate the emerging consequences of using total strengths rather than excess strengths when computing the weighted assortativity coefficient.
To this end, let $\tilde{\rho} = r^\omega_{(\alpha = 1, \beta = 1)}$ be the generalised assortivity coefficient as defined in Eq.\,(\ref{eq:r_new}) which is based on excess strengths and $\rho$ be the assortativity coefficient based on total strengths as in\,\cite{Yuan}.
For the given network, the generalised assortativity coefficients for the different modes of assortativity are $\tilde{\rho}_{\textit{out-in}} = -0.65, \tilde{\rho}_{\textit{out-out}} = -0.76, \tilde{\rho}_{\textit{in-in}} = -0.70$ and $\tilde{\rho}_{\textit{in-out}} = -0.82$.
On the contrary, the coefficients based on total strengths are $\rho_{\textit{out-in}} = 0.29, \rho_{\textit{out-out}} = -0.29, \rho_{\textit{in-in}} = -0.56$ and $\rho_{\textit{in-out}} = -0.82$.

\cite{Yuan}, therefore, conclude that the example network simultaneously shows assortative and disassortative mixing, whereas, as a matter of fact, the network shows no assortative tendencies at all. The network is purely disassortative.
Except for the \textit{in-out} mode of assortativity, the resulting coefficients based on total strengths are throughout greater compared to those of the generalised coefficient.
As mentioned before, for the \textit{in-out} mode of assortativity the excess in- and out-strengths equal the total in- and out-strengths of the ends of an edge, respectively, and thus, $\tilde{\rho}$ and $\rho$ coincidentally exhibit the same value.
Other than that, using total strengths rather than excess strengths will lead to an overestimation of the assortivity towards the assortative direction. The reason for this is that the vertex values of two vertices are artificially inflated by the weight of the connecting edge if total strengths are used.

In order to show this, consider the relative difference $d$ between two variables $x \geq 0$ and $y \geq 0$ defined as $ d = \frac{|x-y|}{(x+y)}$, for which we set $d=0$ if $x = y = 0$.
Computing $d$ with respect to the vertex values of a network yields an indicator of the magnitude of assortativity of a particular connection.
To be precise, if $d$ is small, within its bounds $[0,1]$, then the considered vertices are similar with respect to their vertex values, vice versa, if $d$ is large, then the considered vertices are different with respect to their vertex values, which basically equals the definition of assortative mixing.

Focusing on the \textit{out-in} mode of assortativity, for the edges of the network in Fig.\,\ref{fig:sample_weighted_network} we obtain the following relative differences with respect to excess (out- and in-) strengths $d_{(\tilde{s}^\textit{out},\tilde{s}^\textit{in})} = \{0.33, 1, 1, 1, 1, 1, 1\}$. It can be seen that edges $2$ to $7$ are correctly identified as disassortative.
They all tie a connected vertex to a vertex that is not connected at all, which is the most disassortative connection one can think of.
The edge connecting vertices A and B is identified as rather assortative, since its value of $d_{(\tilde{s}^\textit{out},\tilde{s}^\textit{in}), 1} = 0.33$ is rather small.
By comparing this to the relative differences with respect to total (out- and in-) strengths $d_{(s^{\textit{out}}, s^{\textit{in}})} = \{0.1,0.63,0.38,0.29,0.5,0.2,0.45\}$, which is possible because both $d_{(\tilde{s}^\textit{out},\tilde{s}^\textit{in})}$ and $d_{(s^{\textit{out}}, s^{\textit{in}})}$ are unitless and of the same scale, it can be seen that every single edge is considered more assortative than it actually is, since $d_{(s^{\textit{out}}, s^{\textit{in}}),i} < d_{(\tilde{s}^\textit{out},\tilde{s}^\textit{in}),i}$ for all edges $i$.
This explains why the results $ \tilde{\rho}_{\textit{out-in}} $ and $ \rho_{\textit{out-in}} $ differ so drastically.

\section*{Acknowledgments}

The authors thank the participants of the $ 7^{\text{th}} $ International Conference on Complex Networks and Their Applications (Complex Networks 2018) for helpful discussions and comments.


\end{document}